\documentclass[review]{elsarticle}

\usepackage{lineno,hyperref}
\modulolinenumbers[5]

\journal{Arxive}

\usepackage{amsmath}
\usepackage{amsfonts} 
\usepackage{bbm}
\usepackage{bm}
\usepackage{xcolor}
\usepackage[ruled, vlined]{algorithm2e}
\usepackage{subcaption}
\usepackage{graphicx}
\usepackage{multirow}
\usepackage{array}
\usepackage{booktabs}
\usepackage{tabularx}

\DeclareMathOperator*{\argmin}{arg\,min}

\begin{document}

\begin{frontmatter}
\title{Development of a Two-Level ML Spatial-temporal Framework for Industrial Thermal Striping Applications}
\author[MIT]{Yu-Jou Wang}
\author[MIT]{Emilio Baglietto}
\author[MIT]{Koroush Shirvan}
\address[MIT]{Department of Nuclear Science and Engineering, Massachusetts Institute of Technology, Cambridge, MA 02139, USA}

\begin{abstract}
A data-driven framework for spatial-temporal prediction is proposed for reducing the computational cost of industrial thermal striping applications. The framework aims to efficiently identify the flow features and utilize them in spatiotemporal field emulations with a limited number of full-order simulations. In a parameterized system, the classical projection-based surrogates often suffer from Kolmogorov-width limitation and have challenges with reducibility in highly-nonlinear systems. A two-level machine learning framework is proposed to address these issues, based on physical flow structures. The selection of machine learning algorithms and information extraction is enabled by the idea that the thermal striping phenomenon is driven by large turbulent coherent flow structures. In the first level, the turbulence coherent structures are identified and collected by performing proper orthogonal decomposition on local parameters. A tree-based machine-learning model is then used to down-select the reference structures based on the physical meaning of the modes. In the second level, the reference structure information is broken down into pointwise information for training a deep structure corrector, which corrects the bias between the locally true and reference structures. Demonstration of the framework on triple jet striping test shows that the method can capture the fluctuation frequencies and amplitudes of the spatiotemporal fields in a highly nonlinear setting.  The result shows that the prediction root-mean-squared error of velocity is 0.164 (m/s) for all 2304 spatial points. 
\end{abstract}
\end{frontmatter}

\section{Introduction}
Thermal striping is recognized as a potential cause of high-cycle thermal fatigue damage in many flow engineering problems. 
The damage can arise at locations where the turbulence mixing of different temperature flows drives the spatial temperature gradients and cyclic thermal loads. 
Physics-based modeling and simulation are essential for evaluating this phenomenon, as its severity cannot be assessed via plant instrumentation due to the high frequencies involved, as well as the complex interaction between the source of the ﬂow oscillation and the affected locations. 
Evaluating this phenomenon often requires very high computational cost, as high flow resolution is required to capture the spatial-temporal fluctuations of a thermal field. 
In order to comprehensively capture the underlying physics without over-reliance on empirically driven models and parameters, surrogate modeling of high-fidelity physics tools is commonly pursued. 
A successful development of surrogates can accelerate deployment of new designs and transform operational and maintenance activities, within which multi-query simulations are often required. 
Depending on the construction process, surrogates can be classified into (1) intrusive and (2) non-intrusive approaches. 
The former refers to approaches that require access to the high-fidelity system of equations, often achieved by Galerkin projection\cite{bennerModelReductionApproximation2017}. 
On the other hand, the latter is data-driven and does not require modifications of system solvers. 
Non-intrusive approaches provide versatility toward industrial adoption, as they can readily address extremely complex geometries and large data processing from high-fidelity physics-based tools and experimental measurements.  

Here we focus the discussions on non-intrusive approaches for parameterized systems. 
Surrogates for reducing computational complexity can generally be classified into three categories: (1) Reduced physics, (2) Surrogate functions, and (3) Reduced dimensionality. 
In the first category, only the dominant physics is considered, and the models seek to approximate the solutions through homogenization or averaging. 
Classic examples include one-dimensional lumped parameter models and Reynolds-Averaged Navier-Stoke (RANS) equations. 
Recently, variants combined with ML algorithms have gained increased attention. 
Improvements in correlations development using ML have been seen, such as the development of drag coefficient\cite{heSupervisedMachineLearning2019,jaffarPredictionDragForce2020}, interfacial thermal resistance\cite{wuPredictingInterfacialThermal2019}, and critical heat flux\cite{parkWallTemperaturePrediction2020,zhaoPredictionCriticalHeat2020}. Turbulence closure learnings with various physics-informed techniques have also witnessed growing interest\cite{jiangInterpretableFrameworkDatadriven2021,lingReynoldsAveragedTurbulence2016,wangPhysicsinformedMachineLearning2017}. 
The second category, surrogate functions, aims at finding the direct mapping between input parameters and desired figures of the merit of a system. 
The system's underlying physics is treated as unknown or neglected; the surrogate functions are solely tasked with emulating the input-output behaviors. 
ML algorithms are especially advantageous for this purpose. 
Various schemes and learning targets applied to fluids are proposed in diverse applications\cite{bruntonMachineLearningFluid2020,changClassificationMachineLearning2019,}. 
The last category, dimensionality reduction, transforms and approximates the solutions from a high-dimensional space to a low-dimensional space. There is a rich history of searching for optimized low-dimensional basis (i.e., manifolds). This approach can be both intrusive and non-intrusive\cite{chenGreedyNonintrusiveReduced2018,xiaoNonintrusiveReducedOrder2015,xiaoNonintrusiveReducedorderModelling2015,yuNonintrusiveReducedorderModeling2019}. 
Projection-based reduced-order models using linear manifolds have demonstrated their power in some fluid applications. The trend is also rising in using machine learning algorithms to extract nonlinear manifolds, i.e., manifold learning [13,21,22]\cite{bruntonMachineLearningFluid2020,gonzalezDeepConvolutionalRecurrent2018,leeModelReductionDynamical2020}4.

Developing surrogates for predicting spatial-temporal fluctuations (STF) in complex systems remains challenging. Firstly, in the context of reduced physics models, it is found that the reduced physics models, e.g., Unsteady RANS (URANS, perform poorly in predicting field fluctuations\cite{shamsSynthesisCFDBenchmarking2018}, and the power spectrum can be seriously underestimated\cite{fengDemonstrationSTRUCTTurbulence2020}. 
Attempts to use ML-aided low-to-high-fidelity bias learning could also be challenging. 
Due to the intrinsic homogenizing (averaging) formulation of URANS, only very large-scale turbulence structures can be captured, which prevents the) URANS model from observing the field oscillations\cite{wangHighFidelityDigital2022}.
 On the other hand, Closure learnings often suffer from the issues of generality and physics validity, as many ML architectures do not readily incorporate physics constraints, such as invariance and global conservation laws. 
 The second type, finding the direct spatial-temporal mapping using surrogate functions, may also fail in the presence of the STF's high-dimensional characteristics (i.e., responses covering muti-spatial locations and multi-temporal steps). 
 Although deep-learning algorithms such as Convolutional Neural Networks (CNN) or Long Short-Term Memory (LSTM) could be applied, they are usually very data-hungry. 
 This essentially increases computation cost and limits its application to simple geometries and low computational complexity problems. 
 The third type, dimensional reduction techniques, has the merit of extracting insightful and structurally organized spatial-temporal information from data. 
 However, it is known that the linear manifolds extracted from the projection-based method, such as proper orthogonal decomposition (POD), often suffer from Kolmogorov n-width limitations and have limited reducibility for highly-nonlinear problems. 
 Unfortunately, methods of searching for nonlinear manifolds using machine learning also require extensive high-fidelity data for training. 

Surrogates for STF for industrial applications are in such dilemma: high-fidelity simulations are required to observe the oscillations; the nonlinearity of the spatiotemporal response cannot be easily captured by linear methods; the nonlinear methods, on the other hand, often require a large number of full-order simulations, which extensively increase the computational costs. 

In this research, we focus on developing a robust data-driven surrogate for spatiotemporal fluctuations in industrial thermal striping applications. Thermal striping is a common phenomenon involving spatial-temporal dynamics in industrial applications. The undesirable fluid-structure interaction caused by thermal striping is a crucial phenomenon impacting the reliability of industrial components, such as mixing plenums, junctions, and heating pipes \cite{dahlbergDevelopmentEuropeanProcedure2007,kamayaAssessmentThermalFatigue2014,blondetHighCycleThermal2009}. 
These fluctuations caused by incomplete mixing can potentially cause cyclic thermal stresses on the pipe wall, leading to fatigue cracking. 
The goal of the surrogate is to emulate fluctuation behavior under quasi-steady-state conditions, where the mean flow field remains unchanged, while the instantaneous signals continue to fluctuate. 
Our approach differs from general reduced order modeling techniques, which typically evaluate the system response using known initial conditions. 

The article is structured as follows: Section \ref{sec:problem}. presents the general formulation of a parameterized Navier-Stokes equation; 
Section~\ref{sec:lpod} introduces a technique for low-dimensional manifold extraction using local POD strategy; the framework of constructing a spatiotemporal signal emulator is introduced in Section~\ref{sec:m1} $\sim$ Section~\ref{sec:corrector}. Section~\ref{sec:demo} presents a demonstration of the proposed framework: finally, conclusions are drawn in Section~\ref{sec:conclusion}.
\newcommand{\uvec}{\bm{u}}
\newcommand{\xvec}{\bm{x}}
\newcommand{\muvec}{\bm{\mu}}
\newcommand{\mucase}[1]{\bm{\mu}^{(#1)}}
\newcommand{\phivec}{\bm{\phi}}
\newcommand{\evec}{\bm{e}}
\newcommand{\epsilonvec}{\bm{\epsilon}}
\newcommand{\Pspace}{\mathcal{P}}
\newcommand{\lib}{\bm{L}}
\newcommand{\Dm}[1]{\bm{D}_m(#1)}
\newcommand{\Dmhat}[1]{\hat{\bm{D}}_m(#1)}

\
\section{A two-level Machine Learning Framework for non-intrusive reduced order modeling}
\label{sec:method}

\subsection{Problem Formulation}
\label{sec:problem}
The parameterized Navier-Stokes equations can be written as the following conservative form:
\begin{align}
    \frac{\partial \uvec(\xvec, t; \muvec)}{\partial t} 
    + \nabla \cdot \bm{f}( \uvec(\xvec, t; \muvec); \muvec) 
    = \bm{s}( \xvec, t, \uvec(\xvec, t; \muvec); \muvec) 
    & & \forall (\xvec, t, \muvec) \in \Omega \times  \mathcal{T} \times \Pspace \nonumber \\
    \label{eq:ns}
\end{align}
where $\uvec$ is the field observables; $\bm{f}$ and $\bm{s}$ are the flux term and source term, respectively. 
Here $\Pspace \subset \mathbb{R}^d$ denotes the parameter space with $d$ dimensions; 
$\Omega \subset \mathbb{R}^m$ and $\mathcal{T} \subset \mathbb{R}^+$ are the spatial and temporal domains, respectively.

The system described in Eq.~\ref{eq:ns} can be discretized and solved by high-fidelity full-order model simulations. The high-fidelity solution snapshots corresponding to a given parameter vector  $\muvec$ are represented as $\uvec_h(t; \muvec) \in \mathbb{R}^{N_x}$, where $N_x$ denotes the number of discretized grids.

In this work, we assume that the solution can be approximated using the following form:
\begin{align}
    \uvec_h(t; \muvec) \approx \sum_{m=1}^{r} \phivec_m(\muvec) a_m(t; \muvec) 
    & & \forall t \in [0, T] 
    \label{eq:rom}
\end{align}
where $\phivec \in \mathbb{R}^{N_x}$ and $a \in \mathbb{R}$ are the spatial basis modes and the temporal coefficients, respectively. 
The parameter $r$ represents the number modes for approximation. 
%

Notably, both the spatial basis modes and the temporal coefficients are assumed to be parametric-dependent. 
In the context of turbulent flow, the dynamics of the system can often be characterized by a few dominant large-scale coherent structures\cite{bonnetReviewCoherentStructures2001,cantwellOrganizedMotionTurbulent1981,fiedlerCoherentStructuresTurbulent1988}.  
These structures play a crucial role in determining the fluctuations observed in the flow field, and their characteristics, such as size, shape, and location, can be influenced by variations in the system's parameters\cite{popeTurbulentFlows2000}.
Therefore, given the significance of the coherent structures in capturing the low-dimensional dynamics of the turbulence, it is more natural to select, or approximate, the spatial modes and the temporal coefficients based on the idea of turbulence coherent structures.

In this study, the design of information extraction and the ML algorithm selection is developed upon the idea of coherent structures. 
Unlike the traditional framework of the reduced basis (RB) method, which attempts to construct a parametric-independent reduce space, the proposed approach takes a distinct viewpoint. 
Instead of constructing a reduced space, the framework aims at achieving two key objectives: (1) identifying the dominating flow eddies and their spatial-temporal relations in the flow (2) utilizing machine learning algorithms to predict the parametric dependency of the eddies and emulate their motions.

\subsection{Construction of the Basis Library: Local Proper Orthogonal Decomposition}
\label{sec:lpod}
Given a parameter sampling 
$\Pspace_d = \left\{  \mucase{1}, \mucase{2}, \cdots, \mucase{N_p} \right\} \subset \Pspace $,
a collection of high-fidelity solutions is first obtained from full-order simulations.
For each sample $\mucase{k} \in \Pspace_d$, 
the time-history from high-fidelity simulation can be collected as a snapshot matrix: 
\begin{equation}
    \bm{S}^{(k)} = \left[ 
        \uvec_h(t_1; \mucase{k}), 
        \uvec_h(t_2; \mucase{k}), \cdots, 
        \uvec_h(t_{N_t}; \mucase{k}) 
        \right] 
    \in \mathbb{R}^{N_x \times N_t}
\end{equation}

Proper orthogonal decomposition (POD) is a widely adopted technique in reduced order modeling in searching for the optimal low-dimensional manifold. 
In the realm of fluid dynamics, POD is often used for extracting the dominant coherent structures in the flow field. 
Originating from probability theory and introduced to the fluid community by Lumley\cite{lumleyStructureInhomogeneousTurbulent1967,lumleyCoherentStructuresTurbulence1981}, the identification of coherent structures using POD has been extensively studied in the literature\cite{bidanFilmCoolingJetsAnalyzed2013,kalpaklivesterPODAnalysisTurbulent2015,meyerTurbulentJetCrossflow2007,wuProperOrthogonalDecomposition2019} 

The POD technique approximates the snapshot matrix using singular value decomposition (SVD):
\begin{equation}
    \bm{S}^{(k)} \approx \bm{U}^{(k)} \bm{\Sigma}^{(k)} \bm{V}^{(k)T}
    \label{eq:svd}
\end{equation}
here $\bm{U}^{(k)} \in \mathbb{R}^{N_x \times r}$ and $\bm{V}^{(k)} \in \mathbb{R}^{N_t \times r}$ are the left and right singular vectors, respectively. The diagonal matrix $\bm{\Sigma}^{(k)} \in \mathbb{R}^{r \times r}$ contains the singular values.
The column vectors of $\bm{U}^{(k)}$ correspond to spatial modes, denoted as $\phivec_m(\mucase{k})$, while the row vectors of $\bm{V}^{(k)}$ correspond to temporal coefficients, denoted as $\bm{a}_m(t; \mucase{k})$, in Eq.~\ref{eq:rom}.
The extracted POD modes provide an optimal low-rank approximiation of the high-dimensional spatiotemporal matrix $\bm{S}^{(k)}$.

The above procedure is referred as the \emph{local POD (LPOD)}method, which is different from the traditional global POD method. In LPOD, POD is performed on a local parameter sample, enabling the identification of coherent structures exhibiting similar motion of patterns. In this sense, the POD can be viewed as a tool of flow diagnosis, which decomposes the stochastic turbulence signals into well-organized flow structures. 
The spatial modes give information about where in the field the signal exhibits strong variations, and the temporal correlations give information about how the mode behaves. 

On a contrary, the global POD (GPOD) method involves performing POD on the global snapshot matrix $\bm{S} = \left[ \bm{S}^{(1)}, \bm{S}^{(2)}, \cdots,  \bm{S}^{(N_p)} \right]$. 
Although a globally optimal manifold can be found by GPOD, the resulting modes may not necessarily serve as the most efficient basis for the local parametric space. Particularly in systems exhibiting a high degree of parametric variability, GPOD modes may struggle to capture the local parametric dependencies adequately. It often requires a large number of modes to achieve a reasonable accuracy in the local parametric space.
Thermal striping is a typical example of this type of system. Significant flow pattern variation is often observed in mixing tees, coaxial mixers, and parallel jets(see Fig.~\ref{fig: identified coherent structures}). The nonlinearity of local parametric dependency is often difficult to capture by global POD.

The library is constructed by collecting the POD modes-coefficients-singualr triplets:
\begin{align}
    &\lib = \left\{ 
        L^{(k)}: \forall \mucase{k} \in \Pspace_d
    \right\}\\
    &L^{(k)} = \left\{ 
        \left( \phivec_m(\mucase{k}), \bm{a}_m(\mucase{k}), \sigma_m(\mucase{k}) \right) : \forall m\in 
        \left\{ 1, 2, \cdots, r \right\} 
    \right\}
    \label{eq:lib_k}
\end{align}
    
\subsection{Selection of reference structures}
\label{sec:m1}
In this research, we adopt machine learning to construct a structure predictor, which maps the input parameters to the most probable structures in the library. That is, given a parametric configuration ($\mucase{k}$) and a pre-determined number of modes ($r$), the machine selects $r$ structures with the most resembled descriptors from the library. 
As large turbulent coherent flow structures drive the thermal striping phenomenon, the selection process is based on the physical meaning of the modes.
%
%
In the preprocessing stage, the structures in the library are converted into three descriptors to represent the structure characteristics: 

\paragraph{Spatial descriptors ($\rho_m$)}:
The spatial descriptors are used to represent the spatial distribution of the turbulence coherent structures.
This metric is used to evaluate the size of the active fluctuation region, i.e., the size of the spatial domain in which the spatial basis is greater than a tolerance value
\begin{equation}
    \rho_m(\muvec) = \frac{1}{N_x} \sum_{i=1}^{N_x} \mathbbm{1}(\xvec_i; \muvec)
\end{equation}
where $\mathbbm{1}(\xvec_i)$ is the indicator function at spatial location $\xvec_i$, defined as 
\begin{equation}
    \mathbbm{1}(\xvec; \muvec ) = \begin{cases}
        1, & \text{if } \phivec_m(\xvec; \muvec) > \epsilon \\
        0, & \text{otherwise}
    \end{cases}
\end{equation}
here $\epsilon$ is the tolerance value to determine the active region. In this study, $\epsilon=10\times10^6$ is adopted.

\paragraph{Temporal descriptor ($\nu_m$)}:
For each structure, the temporal coefficients are converted into a power spectrum density (PSD). 
The frequency corresponding to the maximum PSD peak is recorded as the dominant mode frequency $\nu_m$.

\paragraph{Energy descriptor ($\sigma_m^*$)}:
the singular values ($\sigma$) from Eq.~\ref{eq:svd} indicate the ranking of the structures based on their contribution to the total energy. 
The energy descriptor is defined as
\begin{equation}
    \sigma_m^*(\muvec) = \frac{\sigma_m(\muvec)}{\sum_{i=1}^r \sigma_i(\muvec)}
\end{equation}

The descriptor 
$\Dm{\muvec} = \left( \hat{\rho}_m(\muvec), \hat{\nu}_m(\muvec), \hat{\sigma}_m^*(\muvec)\right)$ 
are transformed by a min-max normalizer and stored as the labels of the library component: 
\begin{equation}
    \Dm{\muvec} \Leftrightarrow 
        \left( \phivec_m(\mucase{k}), \bm{a}_m(\mucase{k}), \sigma_m(\mucase{k}) \right)  
\end{equation}

\subsubsection{Training a structure predictor}
The structure predictor is tasked with mapping the input parameters to the most probable library structures based on the predicted descriptors. Given a mode index $m$, the structure predictor takes the input parameters $\muvec$ and outputs the predicted descriptors $\Dmhat{\muvec}$:
\begin{equation}
    (m, \muvec) \mapsto GB_m(\muvec) = \Dmhat{\muvec}
    \label{eq:GB}
\end{equation}

The gradient boosting (GB) multi-output regressor is selected for the prediction. 
With the limited number of library structures for training, the tree-based gradient boosting (GB)method performs well in capturing the nonlinearity and is more robust against overfitting compared to other algorithms, such as the neural network family. 
GB is a forward sequential additive model in which each subsequent model attempts to correct the errors of the previous model on subsets of the data. 
Suppose the GB regressor has M stages, at each stage the model updates its prediction from the previous stage based on the weak learners' output. 
The additive regressor is:
\begin{equation}
    f_{m, i}(\muvec) = \sum_{i=1}^M \gamma_i h_{m, i}(\muvec)
\end{equation}
where $h_{m, i}$ is the $i^{th}$ learner; $\gamma_i$ is the corresponding learning rate. The learner is obtained by minimizing the loss function:
\begin{equation}
    h_{m,i}(\mucase{k}) = \argmin_{h_m} \mathcal{L} \left( y(\mucase{k}),  f_{m, i-1}(\mucase{k}) + h_m(\mucase{k}) \right)
\end{equation}
where $y(\mucase{k})$ is the prediction target; $y(\mucase{k}) = \Dm{\mucase{k}}$ in this case. The term $\mathcal{L}$ is the loss function. The least square error is chosen in this study.

In the prediction stage, the machine outputs are compared with the structure descriptors in the library using $L^2$-norm. The structures with the lowest $L^2$-norms are selected.
The reference structures are then selected by finding the most resembling structure in the library:
\begin{equation}
    \begin{split}
    &\phivec_m^{ref}(\muvec)  = \phivec_m(\muvec^*) \\
    &a_m^{ref}(t; \muvec) = a_m(t; \muvec^*) \\
    &\muvec^* = \argmin_{\hat{\muvec}} 
    \lVert
        \bm{D}(m, \muvec) - \bm{D}(m, \hat{\muvec})
    \rVert_2 \\ 
    \end{split}
    \label{eq:lookup}
\end{equation}

\subsection{Structure Correction}
\label{sec:corrector}
After obtaining the reference structures, the next step is to obtain the reduced basis by correcting for the bias introduced by using the reference structures. 
%
Here we define the spatial bias as the difference between the selected library structure and the true structure:
\begin{align}
   \evec_m(\muvec) = \epsilon_m(\muvec, \muvec^*) &=  \phivec_m^{ref}(\muvec) -  \phivec_m(\muvec) \\
    &= \phivec_m(\muvec^*) - \phivec_m(\muvec)
    \label{eq:bias computation}
\end{align}
where $\phivec_m(\muvec)$ is the true spatial mode; $\phivec_m{(\muvec^*)}$ is the reference spatial mode obtained by the structure predictor(as introduced in Sec.~\ref{sec:m1}), corresponding to the $m^{th}$ mode extracted from the snapshot matrix of parameter $\muvec^*$.

\begin{figure}
    \centering
    \includegraphics[width=\textwidth]{ 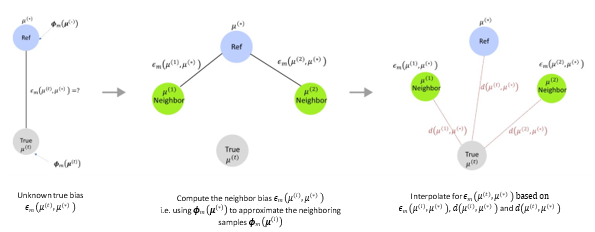}
    \caption{Nonlinear interpolation for the bias.}
    \label{fig:nonlinear_interpolation}
\end{figure}

Without a prior knowledge of the target bias, we propose using nonlinear interpolation to estimate the target bias $\epsilon_m(\muvec, \muvec^*)$. 
The method leverages the information of the neighboring samples to predict the target bias.
The conceptual schematics of the interpolation is illustrated in Figure \ref{fig:nonlinear_interpolation}.
 
Suppose a reference structure $\phivec_m(\muvec^*)$ is obtained from the structure predictor given the target input parameter $\mucase{t}$. 
The bias of using $\phivec_m(\muvec^*) $ to approximate the true structure $\phivec_m(\mucase{t})$ is assumed to be a function of using the same sturcture to approximate the neighboring structures, weighted by the 
parametric distances, which is defined as the Euclidean distance between two parameters:
\begin{equation}
    d(\mucase{a}, \mucase{b}) = \lVert \mucase{a}- \mucase{b} \rVert_2
    \label{eq:parametric distance}
\end{equation}

\begin{equation}
    \epsilon_m(\mucase{t}, \muvec^*) = \mathcal{I} 
        \left(
        \begin{split}
            &w_1 \cdot \underbrace{\epsilon_m(\mucase{n_1}, \muvec^*)}_{\text{neighbor bias 1}} \\
            + &w_2 \cdot \underbrace{\epsilon_m(\mucase{n_2}, \muvec^*)}_{\text{neighbor bias 2}} \\
             &\vdots  \\
            + &w_{n} \cdot \underbrace{\epsilon_m(\mucase{n_n}, \muvec^*)}_{\text{neighbor bias n}} 
        \end{split}
        \right)
\end{equation}
where $w_i$ is the weight of the $i^{th}$ neighbor, which is a function of 
parametric distance $d(\mucase{n_i}, \mucase{t})$; 
$n$ is the number of neighbors; $\mathcal{I}$ is the interpolation function.

\begin{figure}
    \centering
    \includegraphics[width=0.7\textwidth]{ 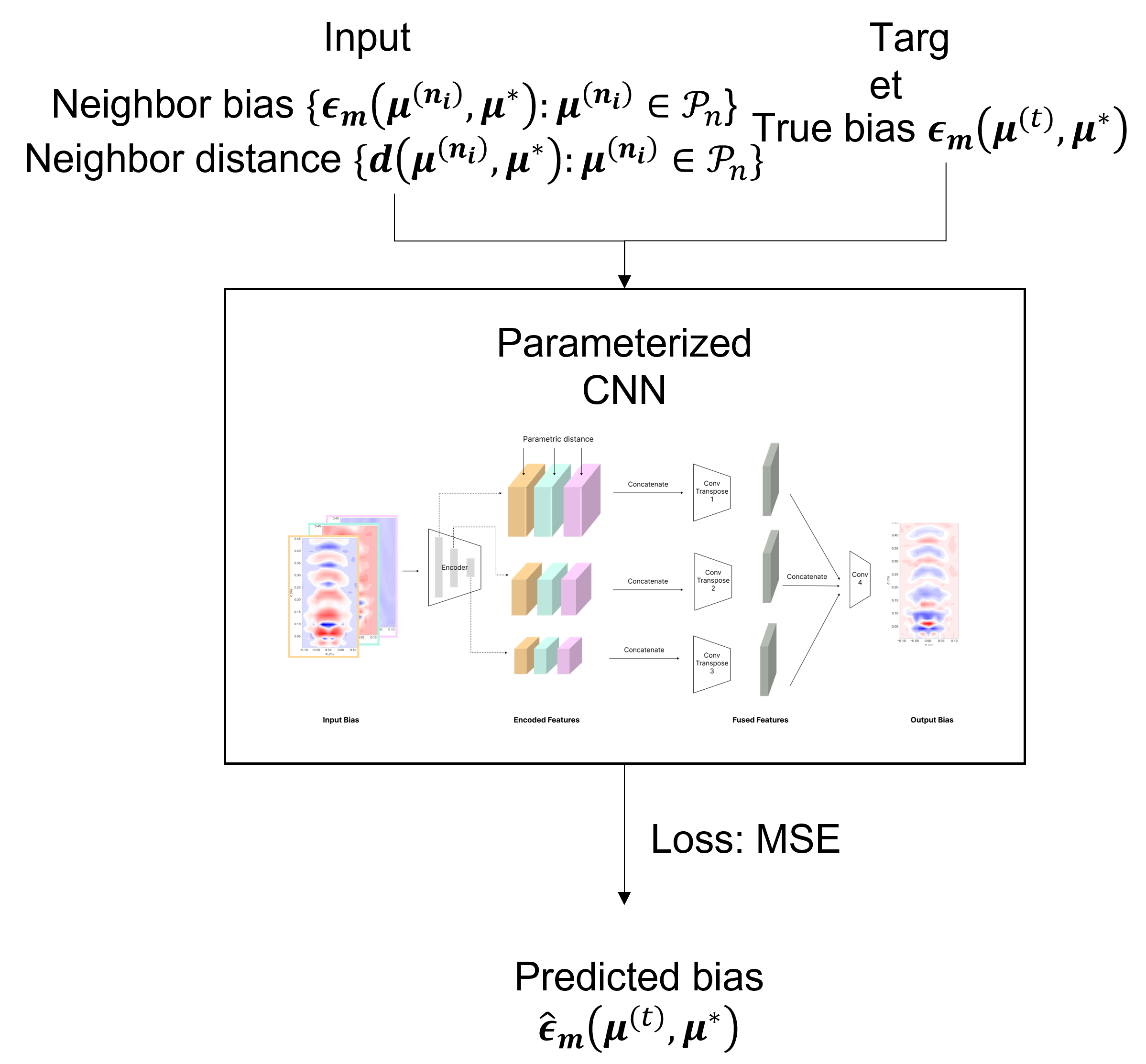}
    \caption{Training workflow of the parameterized CNN}
    \label{fig:training CNN}
\end{figure}

In the proposed framework, we train a parameterized Convolutional Neural Network(CNN) to predict the bias. 
This approach is useful for detecting the deformation and shifting of coherent structures, which is essential for predicting the parametric change of fluid fundamental characteristics. CNNs are well-suited for this task, as they can learn to identify key features and patterns of coherent structures. 
The data preprocessing for the nonlinear interpolation is shown in Algorithm \ref{alg:nonlinear_interpolation}; and the training workflow is illustrated in Figure \ref{fig:training CNN}.









\begin{algorithm}[H]
    \KwIn{number of neighbors considered in interpolation}
    \KwOut{ target bias, neighbor biases, parametric distances
        }
    \For{ $ \forall \mucase{t} \in \Pspace_d$}{
        \For{$m=1$ \KwTo $r$}{
            Compute the library structure $\phivec_m(\muvec^*)$ from Eq.~\ref{eq:lookup}.\\
            Compute the target bias $\epsilon_m(\mucase{t}, \muvec^*) $ using Eq.~\ref{eq:bias computation}.\\
            Select $n$ neighbor samples with shortest parametric distances, forming a neighbor set $\mathcal{P}_n \subset \Pspace_d$.\\
            \For{$i=1$ \KwTo $n$}{
                Collect the neighbor bias $\epsilon_m(\mucase{n_i}, \muvec^*)$ using Eq.~\ref{eq:bias computation}.\\
                Collect the corresponding parametric distance $d(\mucase{n_i}, \mucase{t})$.\\
            }
        }
    } 
    \caption{Data preprocessing for nonlinear interpolation}
    \label{alg:nonlinear_interpolation}
\end{algorithm}

\begin{figure}
    \centering
    \begin{subfigure}{0.7\textwidth}
        \centering
        \includegraphics[width=.8\linewidth]{ 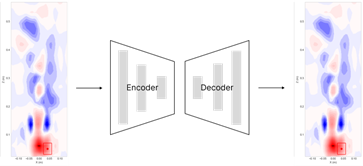}
        \caption{Autoencoder}
        \label{fig:encoder}
    \end{subfigure}
    \begin{subfigure}{\textwidth}
        \centering
        \includegraphics[width=.8\linewidth]{ 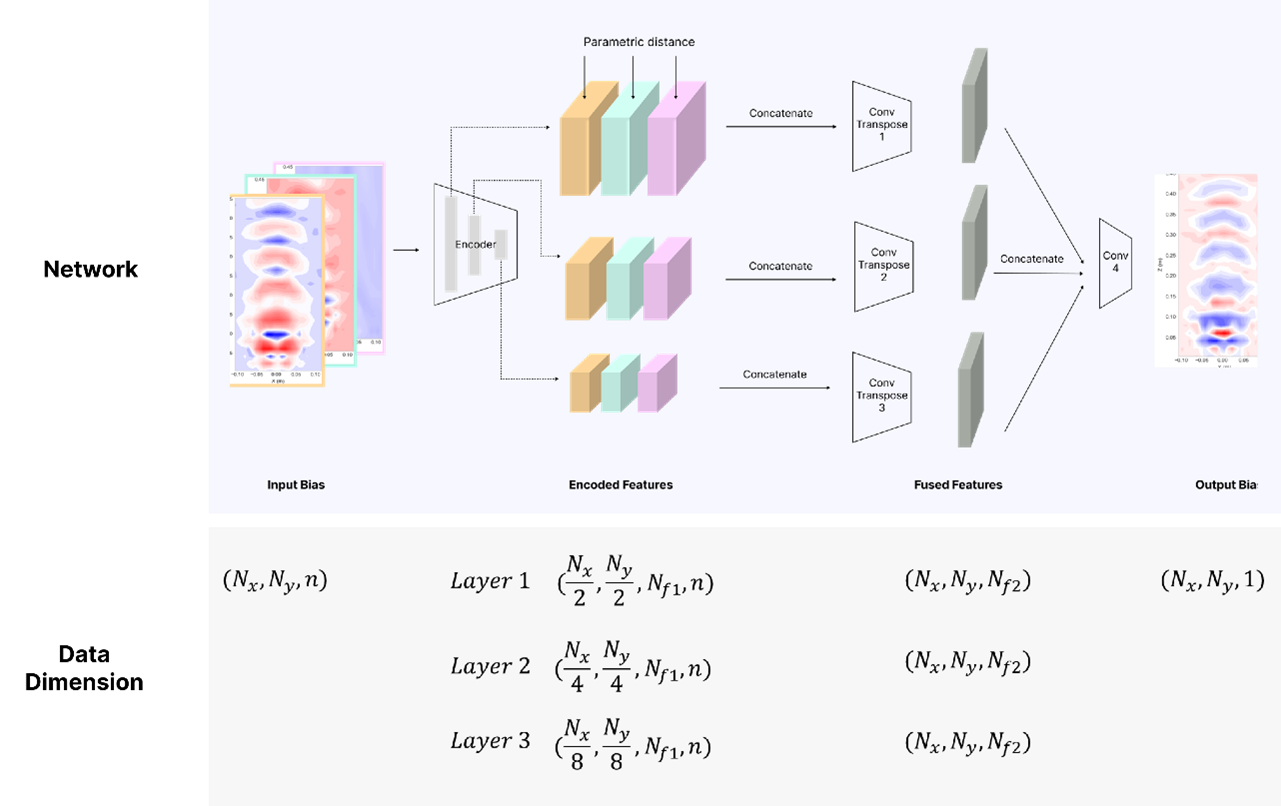}
        \caption{Interpolation}
        \label{fig: CNN interpolation}
    \end{subfigure}
    \caption{Architecture of the parameterized autoencoder model}
    \label{fig:CNN}
\end{figure}

The architecture of the parameterized CNN is shown in Figure~\ref{fig:CNN}.
%
The model is inspired by the feature fusion method proposed by Gong and Yang\cite{gongVideoFrameInterpolation}. 
The model training involves two stages. 
In the first stage, an autoencoder is trained to extract the spatial dependency and the latent features of the 2D bias, 
as illustrated in  Figure~\ref{fig:encoder}.
This is a crucial step in the model's training, as it enables the system to recognize patterns and extract essential information from the data. 
In the second stage, the first three layers of the autoencoder act as a feature extractor. 
During this stage, the encoder’s weights remain fixed, and the n-neighbor biases are individually fed into the encoder. 
After passing through the encoder layers, the features obtained from the first three layers are parameterized with the distance vectors, concatenated, and fused by the following transposed convolutional layers to create a complete representation of the input bias, 
as illustrated in Figure~\ref{fig: CNN interpolation}.
The fused features are then fed into a convolutional layer to perform spatial nonlinear interpolation. 
In this work, the loss function is based on the mean squared error for both stages.

\subsection{Emulation Workflow}
After obtaining the reference structures and the predicted bias, the predicted structure can be computed as:
\begin{equation}
    \hat{\phivec}_m(\muvec) = \phivec_m(\muvec^*) - \hat{\epsilonvec}_m(\muvec, \muvec^*)
\end{equation}
And from Eq.~\ref{eq:rom}, the field variables can be computed by: 

\begin{equation}
    \hat{\uvec} (t; \muvec) = \sum_{m=1}^{r} \hat{\phivec}_m(\muvec) \hat{a}_m(t; \muvec) 
\end{equation}

Here $\hat{a}_m(t; \muvec)$ is the time-dependent coefficients for the structure $m$. In the present study, we set $\hat{a}_m(t; \muvec)$ to be the same as the reference coefficients $a_m(t; \muvec^*)$. 
Although machine learning algorithms exist to predict the time-dependent coefficients, capturing the parametric dependency of coefficient time-histories often requires a large amount of training data.
As will be shown in the numerical demonstration, 
the reference structures selected based on their physical significance,  exhibit highly similar oscillatory behaviors, which justifies our choice of utilizing the reference coefficients for the predicted structures without further transformations.
The exploration and prediction of parametric dependencies in greater detail is left as a potential avenue for future work. 

\begin{figure}
    \centering
    \includegraphics[width=\textwidth]{ 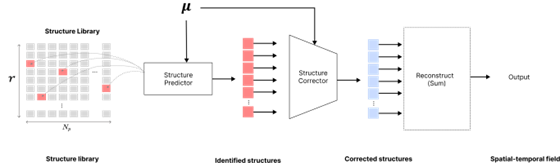}
    \caption{Workflow of the proposed emulation framework.}
    \label{fig:workflow}
\end{figure}

The prediction workflow is illustrated in Figure~\ref{fig:workflow}.

\section{Demonstration}
\label{sec:demo}
\subsection{Experimental Setup}
The triple jet configuration considered in this work is obtained from the experiments of Tokuhiro and Kimura\cite{tokuhiroExperimentalInvestigationThermal1999}. 
Three quasi-planar vertical jets are located at the bottom of the tank. The temperature of the central jet is slightly lower than the two adjacent jets, 
and the velocity of the center (cold) jet, 
denoted by $v_c$, differs from the two adjacent (hot) jets, $v_h$. 
Readers are referred to\cite{tokuhiroExperimentalInvestigationThermal1999} for the detailed geometry. 
The experiment, for which the Reynolds number is $Re_D=1.8 \times 10^4$, 
is commonly used as a benchmark problem to investigate the mixing process of unsteady flow turbulence for industrial applications.

In this study, we perform full three-dimensional numerical simulations with the commercially available software STAR-CCM+ ver.16.02, for high-fidelity training data generation. 
The structure-based hybrid turbulence model, 
STRUCT-$\varepsilon$, is adopted\cite{xuSecondGenerationURANS2020a}, 
having demonstrated the key ability to resolve the coherent unsteady turbulent structures driving the striping phenomenon, at greatly reduced computational cost, while providing Large Eddy Simulation like accuracy. 
Acton et al.\cite{actonStructureBasedResolutionTurbulence2015} compared the STRUCT performance with the LES and experimental data of the Tokuhiro and Kimura triple-jet test case. It was shown that the STRUCT model is capable of predicting the experimental profiles of the time-averaged temperature and velocity, 
the temperature fluctuation and the frequency of temperature fluctuations. 
The numerical practices adopted respect the necessary best practices and mesh requirements to guarantee the required accuracy, 
as previously shown by Feng\cite{fengSTRUCTurebasedURANSSimulations2018}; 
all approximations adopted are second order accurate, where in particular spatial interpolation of the advective terms leverages a non-oscillatory upwind based second-order schemes, while time integration is performed with a second order three time level method\cite{simcenterDocumentationVersion20192019}. 
While the STRUCT-$\varepsilon$ model is selected as the base model for data generation thanks to its proven accuracy and computational efficiency, 
the proposed methodology is model agnostic and can be adopted with any high-fidelity and high-resolution numerical or experimental measurement. 

In this demonstration, we treat $v_c$ and $v_h$ as two independent random variables, 
ranging from $[0.5, 1.5]$ m/s; the velocity ratio $r=  v_c⁄v_h$  will be within the range of $[0.33,3]$.  

The number of training samples are largely constrained by the computational cost of the full-order simulations; the cpu-time for one full-order simulation is 165.6 CPU hours to obtain 15 sec quasi steady-state oscillations. 
In this study, 30 full-order simulations are performed and split into 27 training samples and 3 testing samples. 
The time-history data is collected from the resampled 2D mid-plane in the mixing region, 
which includes 576 discritized points, as illustrated in Figure~{fig:window}. 
In each simulation, 1000 snapshots with time-step equal to 0.005 second are collected.

\begin{figure}[htbp]
    \centering
    \subfloat[Design of experiments]{
        \includegraphics[height = 5cm]{ 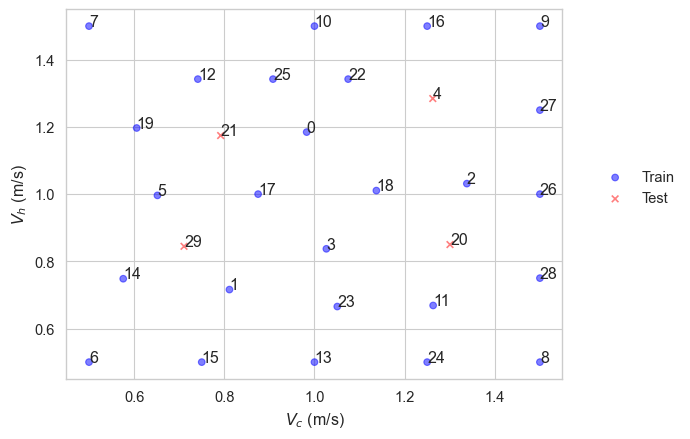}
        \label{fig:doe}
    }
    \hfill
    \subfloat[data window for collecting the snapshots]{
        \includegraphics[height=5cm]{ 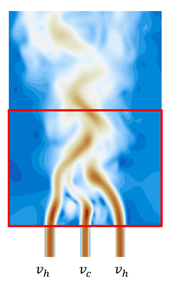}
        \label{fig:window}
    }
    \caption{Experimental setup of the triple jet emulation}
\end{figure}


\subsection{Instantaneous Velocity Field and Coherent Structures Identification}
\begin{figure}
    \begin{subfigure}[b]{0.20\textwidth}
        \includegraphics[width=\textwidth]{ 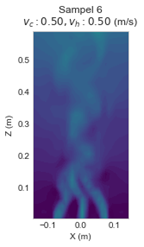}
        \caption{Sample 6}
    \end{subfigure}
    \hspace{\fill}
    \begin{subfigure}[b]{0.20\textwidth}
        \includegraphics[width=\textwidth]{ 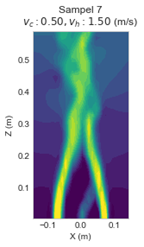}
        \caption{Sample 7}
    \end{subfigure}
    \hspace{\fill}
    \begin{subfigure}[b]{0.20\textwidth}
        \includegraphics[width=\textwidth]{ 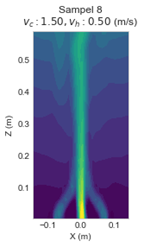}
        \caption{Sample 8}
    \end{subfigure}
    \hspace{\fill}
    \begin{subfigure}[b]{0.20\textwidth}
        \includegraphics[width=\textwidth]{ 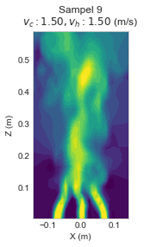}
        \caption{Sample 9}
    \end{subfigure}
    \caption{Instantaneous velocity snapshots of the four cases in the parameter space}
    \label{fig: inst_vel}
\end{figure}

Figure~\ref{fig: inst_vel} shows the instantaneous flow velocity of the four cases in the parameter space, corresponding to the corner values depicted in Figure~\ref{fig:doe}.
As the velocity ratio approaches unity(i.e., the diagonal from the lower left to the upper right), notable oscillations emerge in the flow field; 
as the velocity ratio moves away from unity, the flow stabilizes, and either the central jet (sample 8 ) or adjacent jets (sample 7) becomes dominant.

It is worth noting that flow asymmetry can manifest during the process of turbulence mixing between parallel jets. 
Under specific boundary conditions, the central jet tends to lean towards the adjacent jet on one side, resulting in more pronounced oscillations on that particular side compared to the other (e.g., sample 7). This asymmetry has also been observed in experiments conducted by Kimura et al.\cite{kimuraExperimentalInvestigationTransfer2007}.
Due to the occurrence of jet tilting, capturing the complete symmetry of the jets requires a very long simulation time, which may be impractical for multi-query applications. To ensure data consistency, we address this issue by horizontally flipping the domain, if needed, and manually orienting the tilted snapshots towards the left in cases where the jet oscillations exhibit asymmetry. This approach ensures that the data remains aligned and reliable for subsequent analysis.

\begin{figure}
    \subfloat{
        \includegraphics[width = 3 cm]{ 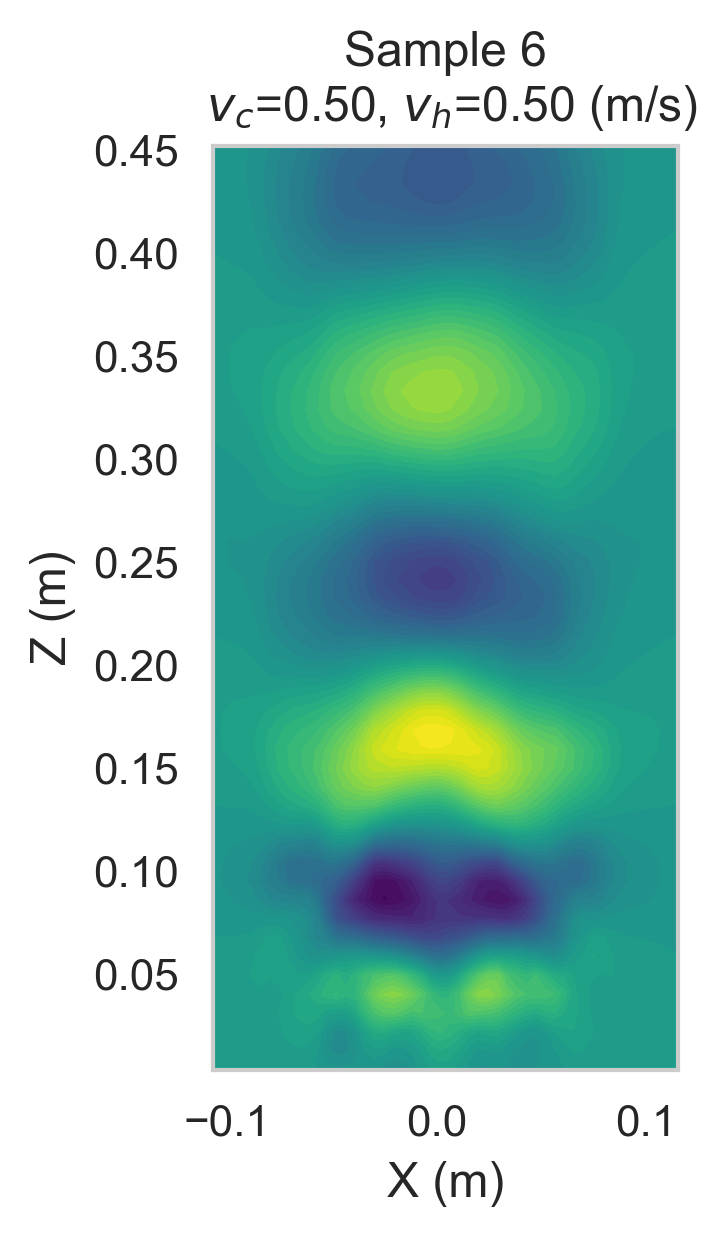}
    }
    \subfloat{
        \includegraphics[width = 3 cm]{ 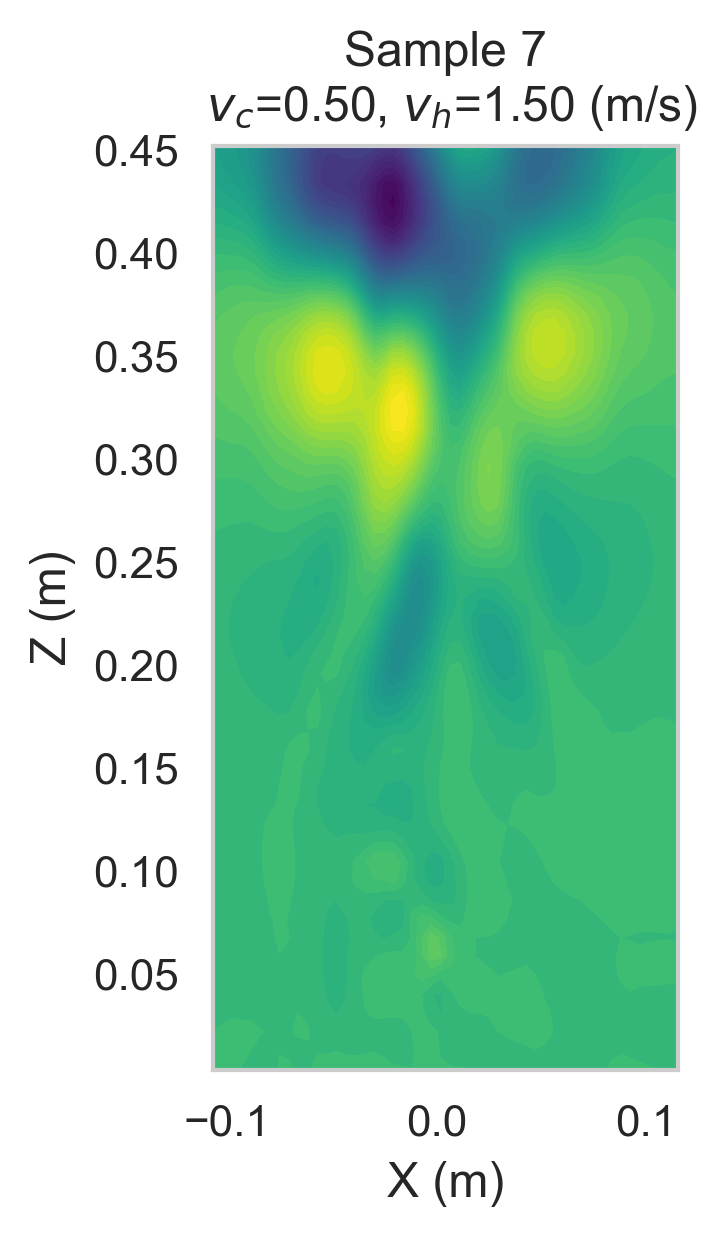}
    }
    \subfloat{
        \includegraphics[width = 3 cm]{ 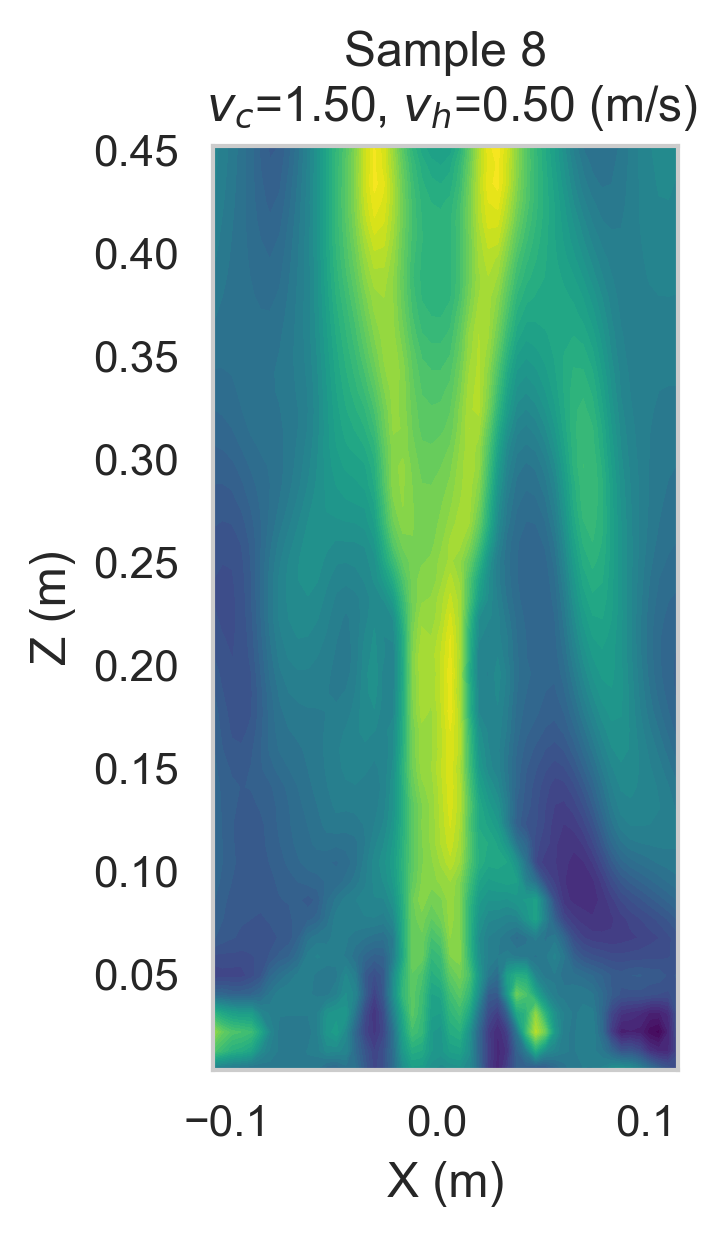}
    }
    \subfloat{
        \includegraphics[width = 3 cm]{ 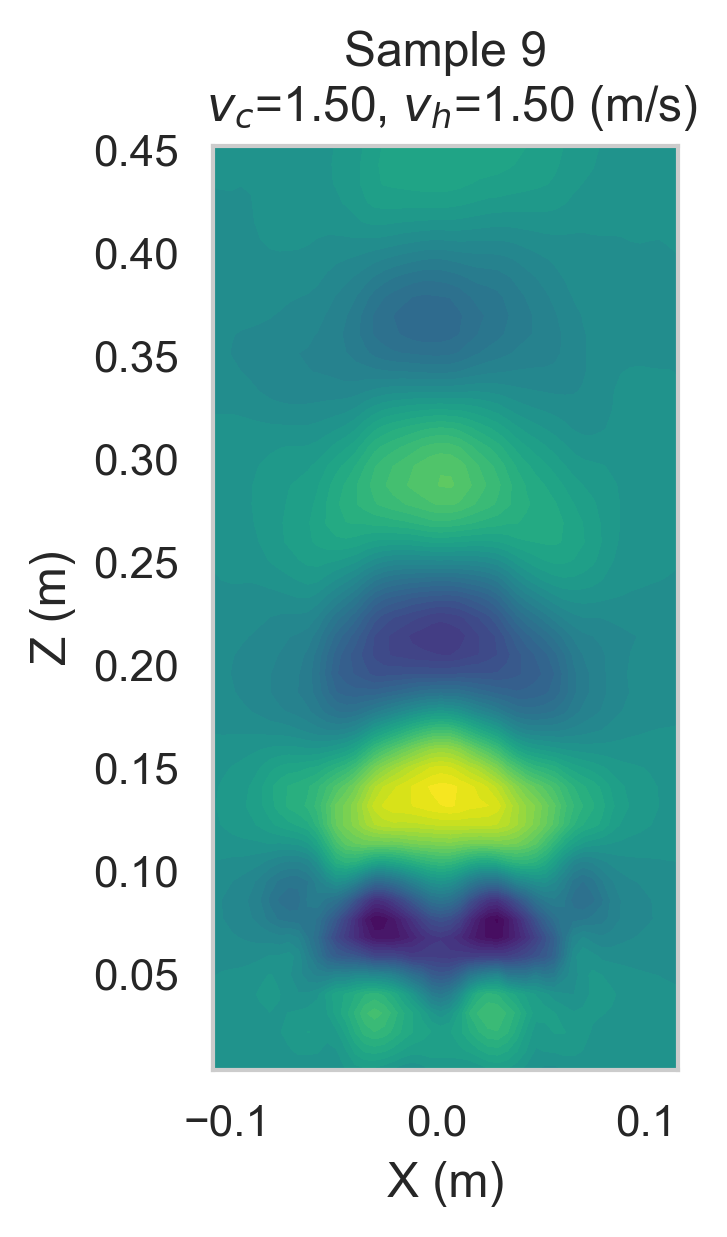}
    }
    \\
    \subfloat{
        \includegraphics[width = 3 cm]{ 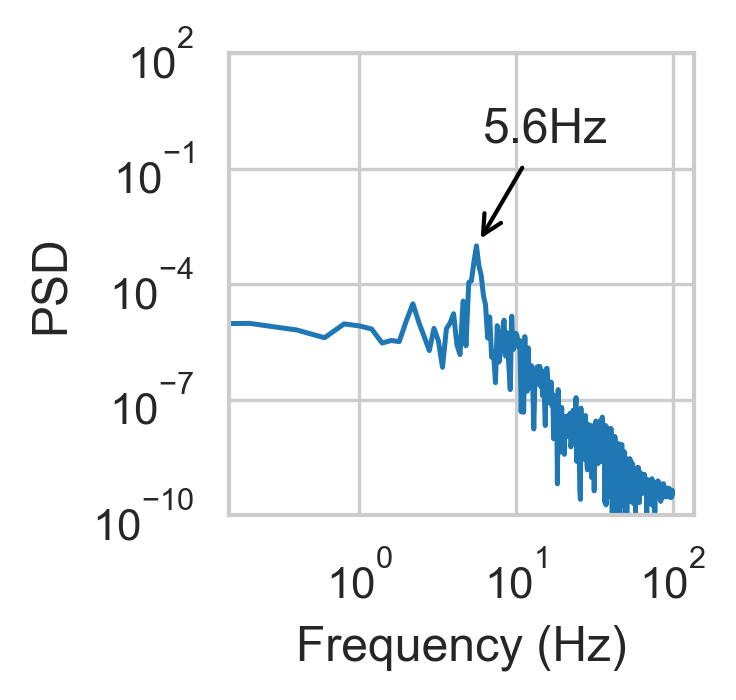}
    }
    \subfloat{
        \includegraphics[width = 3 cm]{ 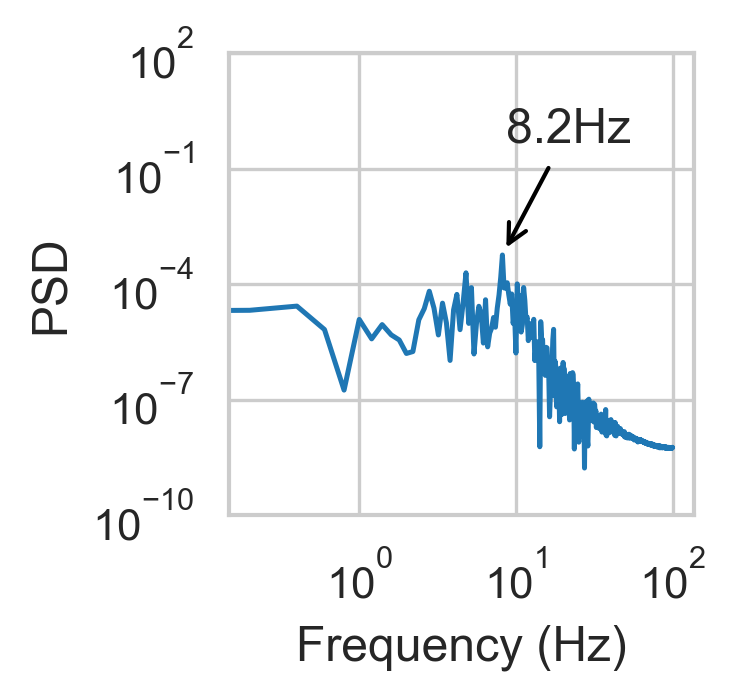}
    }
    \subfloat{
        \includegraphics[width = 3 cm]{ 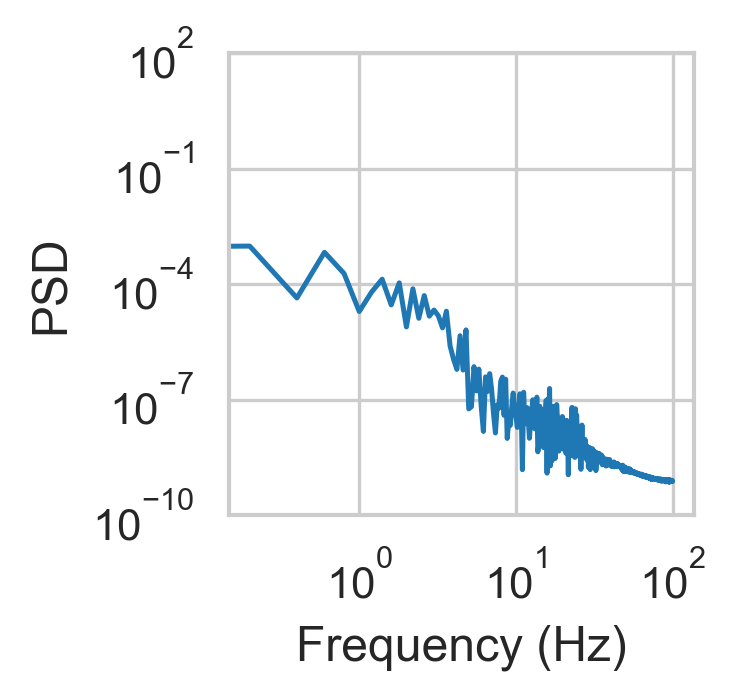}
    }
    \subfloat{
        \includegraphics[width = 3 cm]{ 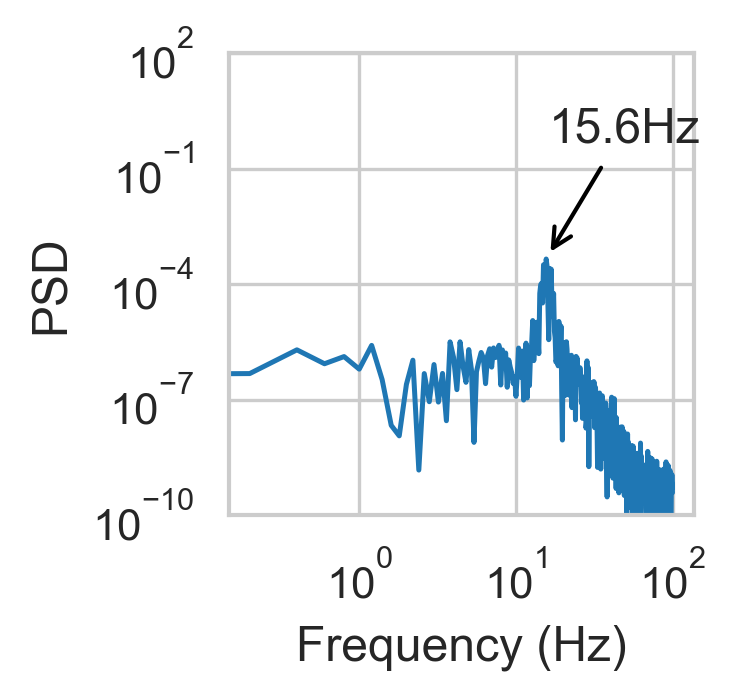}
    }
    \caption{Identified coherent structures and their corresponding temporal coefficients of the four samples. First row: spatial structure; second row:  temporal coefficients(power spectrum density).}
    \label{fig: identified coherent structures}
\end{figure}

Figure~\ref{fig: identified coherent structures} presents the coherent structures and their associated temporal coefficients, represented by the power spectrum density (PSD),  for the four samples. For clarity, only the first mode (highest energy) is displayed, highlighting the dominant oscillatory behavior in the mixing region.
In sample 6 and sample 9, symmetric pairs of structures are observed, indicating the presence of alternating fluctuations in the mixing zone. Notably, strong PSD peaks are observed at frequencies of 5.6 Hz and 15.6 Hz, respectively, corresponding to the dominating oscillation frequencies in these samples. 
In sample 7, the oscillation is primarily observed at locations distant from the injection region ($y>0.25 (m)$), with a slightly less prominent frequency peak at 8.2 Hz. 
In sample 8, the field does not exhibit a clear oscillatory characteristic and no dominant frequency is observed.

\begin{figure}
    \centering
    \includegraphics[width=0.8\textwidth]{ 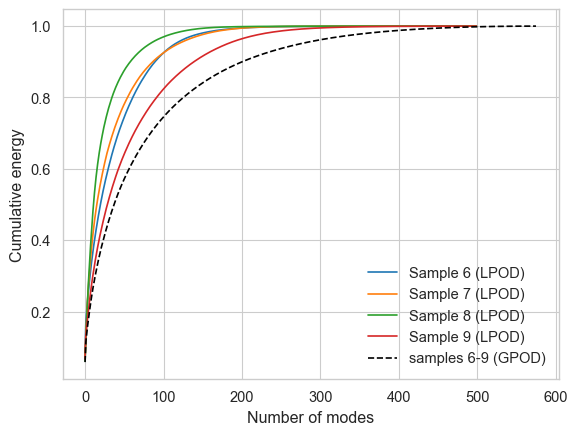}
    \caption{Comparison of energy content(singular values) between local POD(LPOD) and global POD(GPOD)}
    \label{fig: singular values}
\end{figure}

The energy content of the first 600 POD modes for the four samples are shown in Figure~\ref{fig: singular values}. 
With high flow variability, traditional global POD requires a larger number of modes to capture sufficient amount of energy. 

%
%
%
In contrast, the local POD approach, which extracts turbulence coherent structures from local-parametric setting, is more efficient at describing the system. This demonstrates the advantage of using local POD modes to describe the system's response to parameter changes. 
Utilizing local information allows for the capture of the intricate relationships between flow structures and parametric variations, which is valuable in systems with complex flow dynamics.

\subsection{Structure Prediction and Bias Correction}
\begin{figure}
    \centering
    \subfloat[Sample 20]{
        \includegraphics[width = 0.3\textwidth]{ 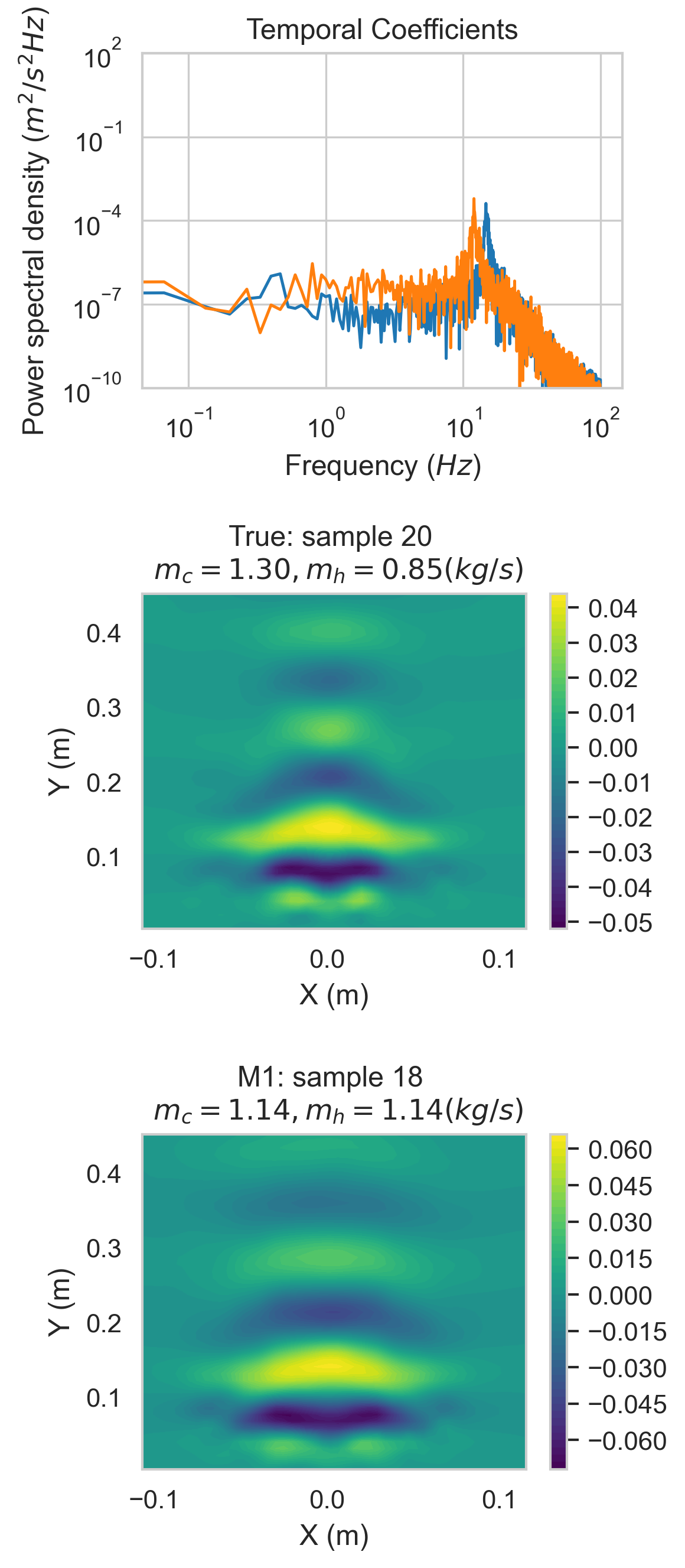}
    }
    \subfloat[Sample 21]{
        \includegraphics[width = 0.3\textwidth]{ 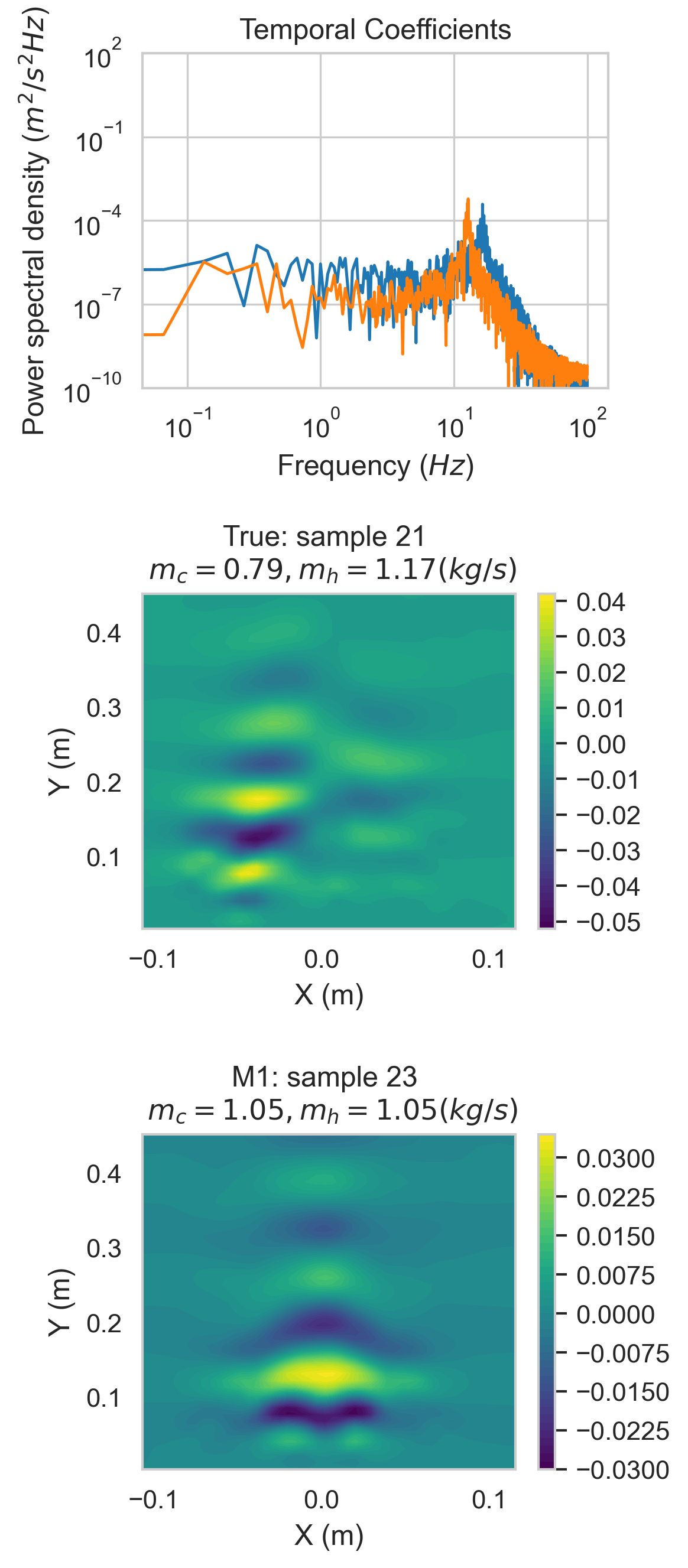}
    }\\
    \subfloat[Sample 4]{
        \includegraphics[width = 0.3\textwidth]{ 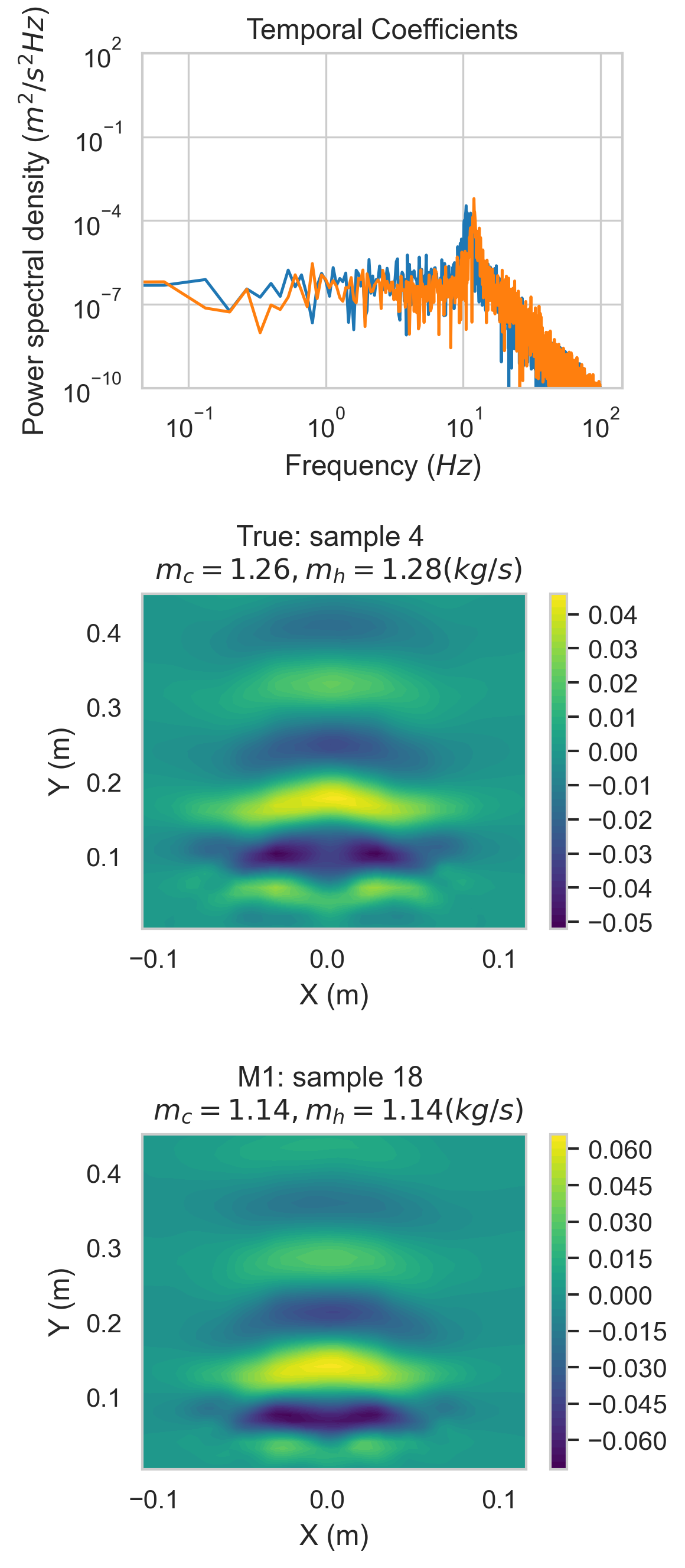}
    }
    \subfloat[Sample 29]{
        \includegraphics[width = 0.3\textwidth]{ 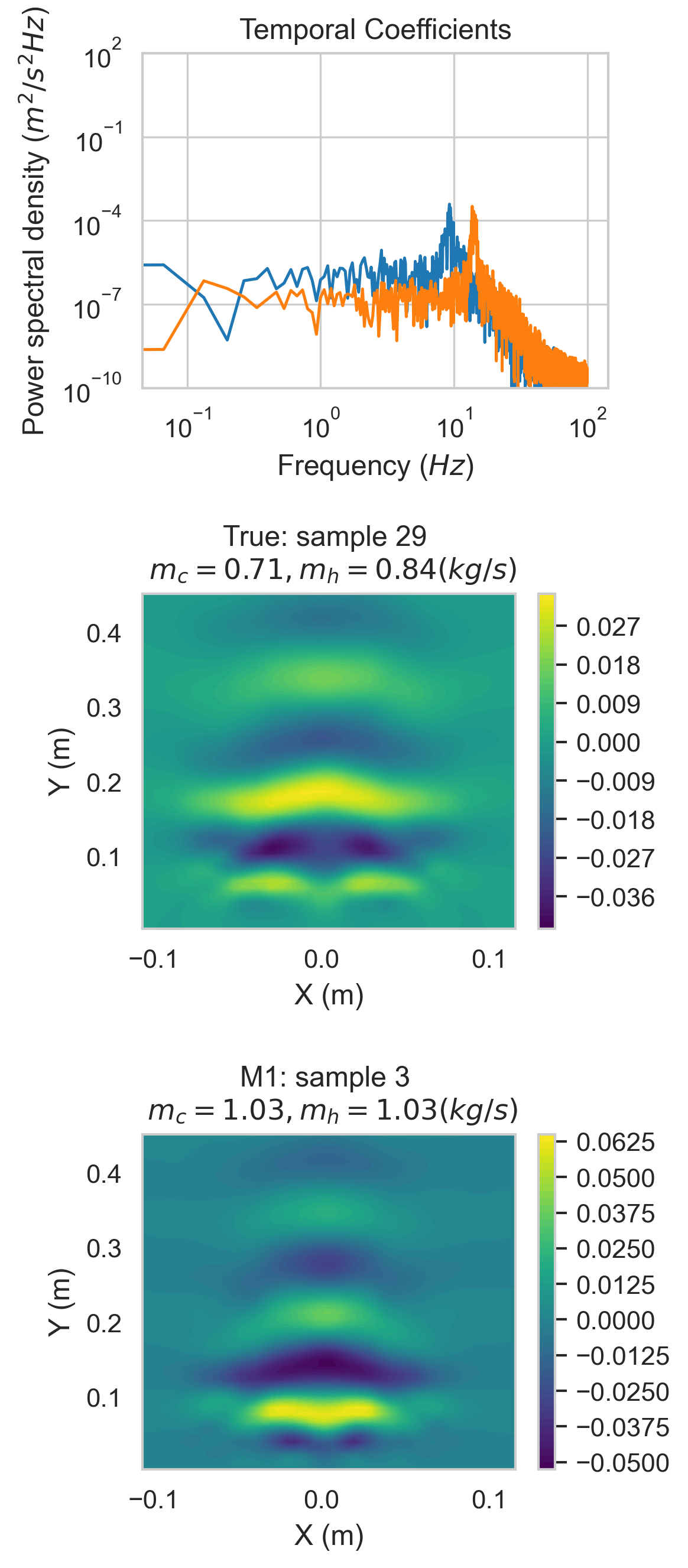}
    }
    \caption{ Performance of structure predictor. The first row is the temporal coefficients of the POD modes. The second and third rows are the true and predicted structures, respectively}
    \label{fig: m1 results}
\end{figure}

In this section, we demonstrate using the structure predictor to find the coherent structures in the parametric setting. A total of 17 samples in the library set are processed through LPOD. The first 20 modes-coefficients-singulars are stored as the library structures. Figure~\ref{fig: m1 results} visualizes the result from the structure predictor (machine 1) of the four testing cases( sample 20, 21, 4, 29). 
Comparing to the true values, the predicted reference structures and the temporal behaviors exhibit very similar characteristics. Notably, the structure predictor can successfully select the structure in the case even when the oscillatory asymmetry is observed.

Although the machine 1 and the structure library can give good reference structures for approximations, the prediction is inevitably subject to bias, as suggested in Figure~\ref{fig: m1 results}. To correct the error introduced by structure difference, we introduce a structure corrector based on local shape bias and nonlinear interpolations. 

\begin{figure}
    \centering
    \subfloat[Sample 20]{
        \includegraphics[width = 0.5\textwidth]{ 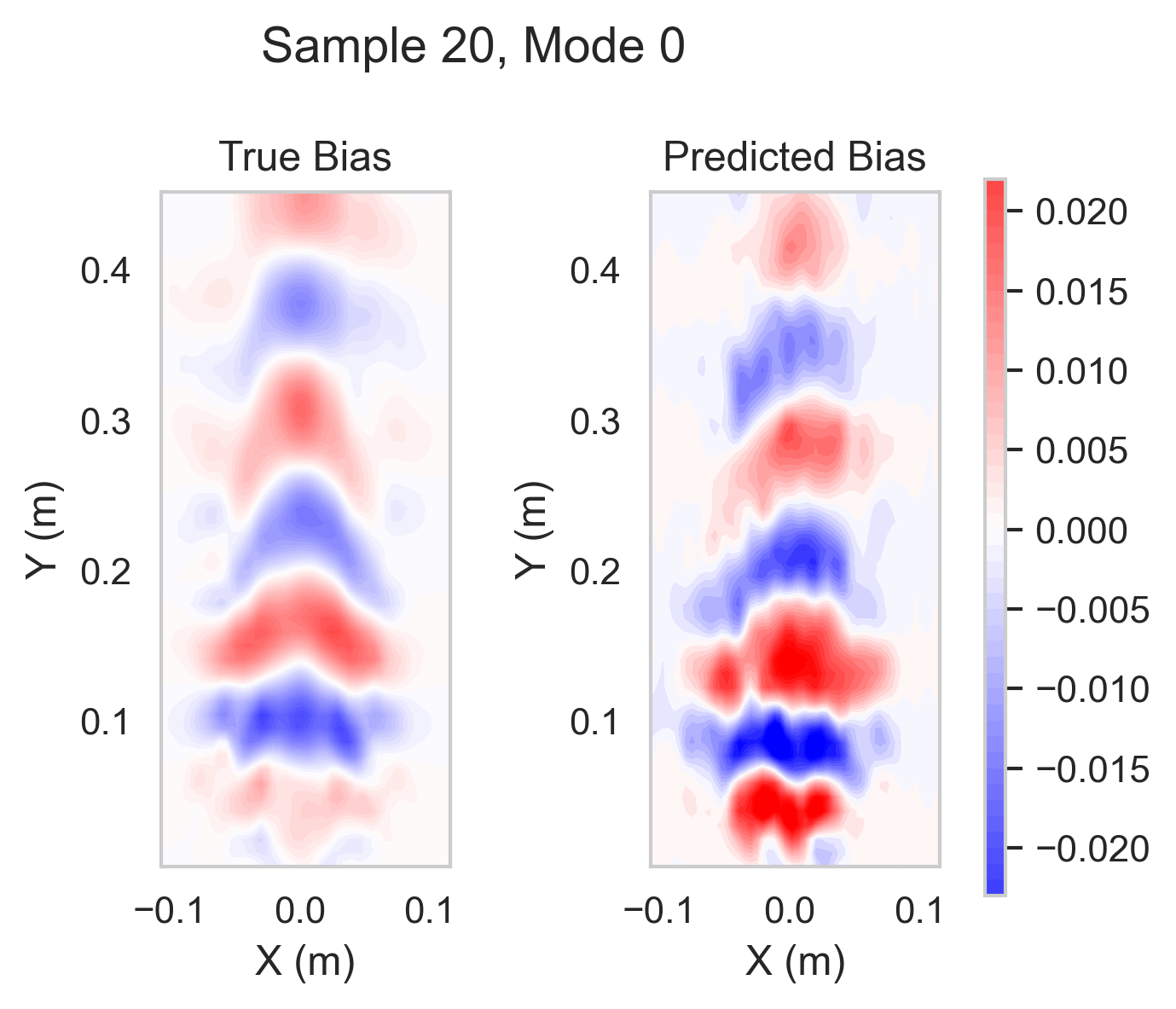}
    }
    \subfloat[Sample 21]{
        \includegraphics[width = 0.5\textwidth]{ 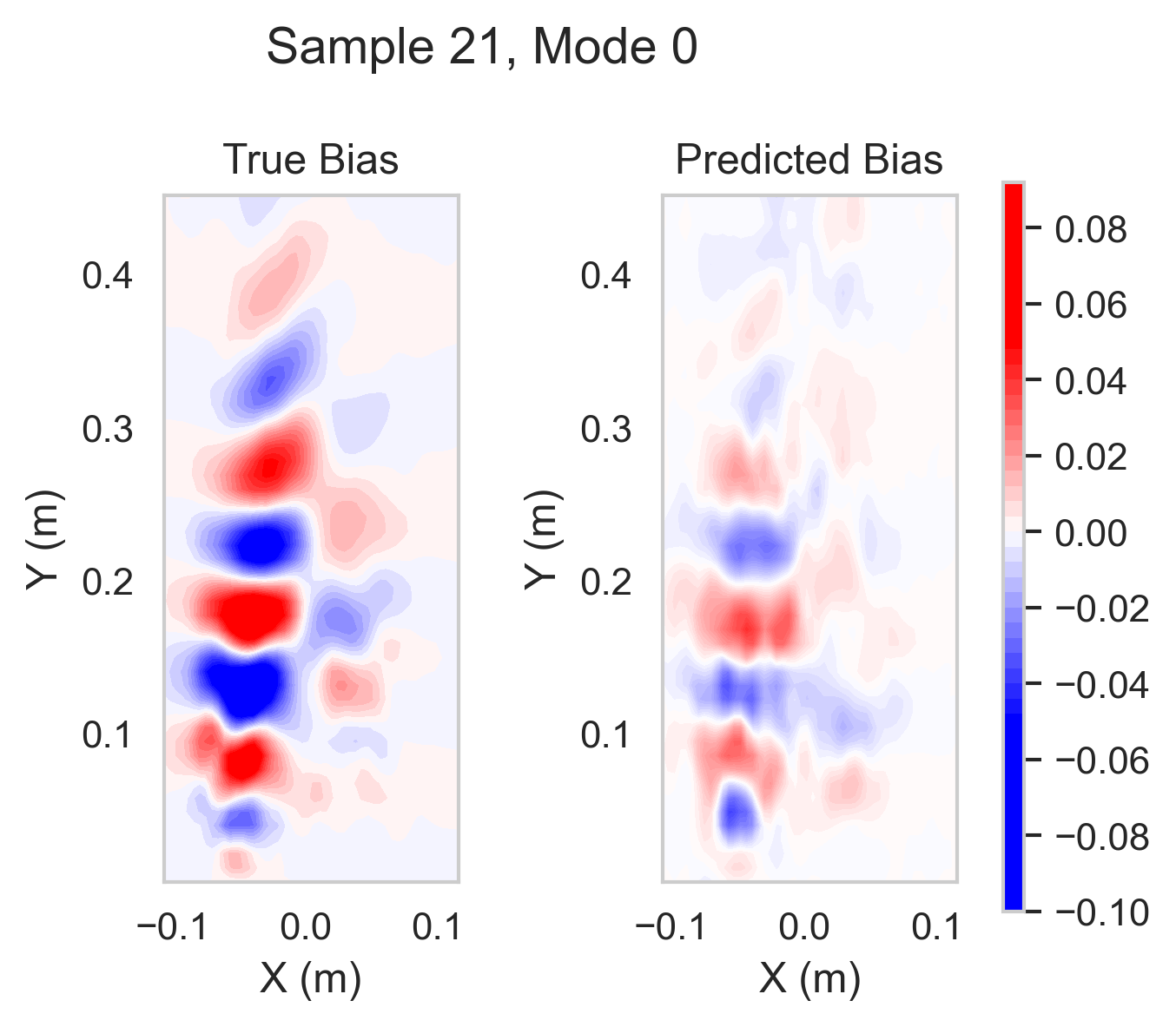}
    }\\
    \subfloat[Sample 4]{
        \includegraphics[width = 0.5\textwidth]{ 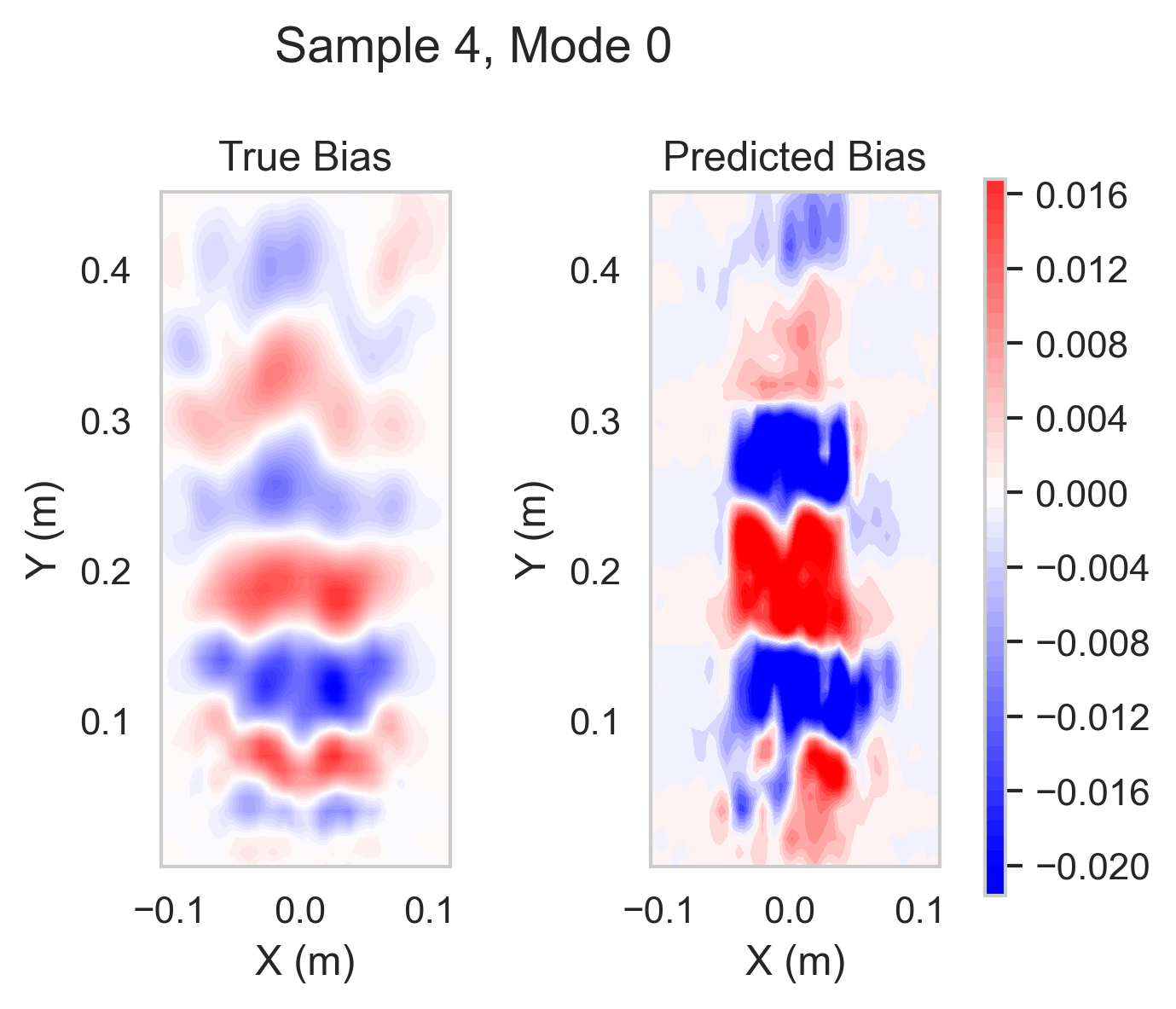}
    }
    \subfloat[Sample 29]{
        \includegraphics[width = 0.5\textwidth]{ 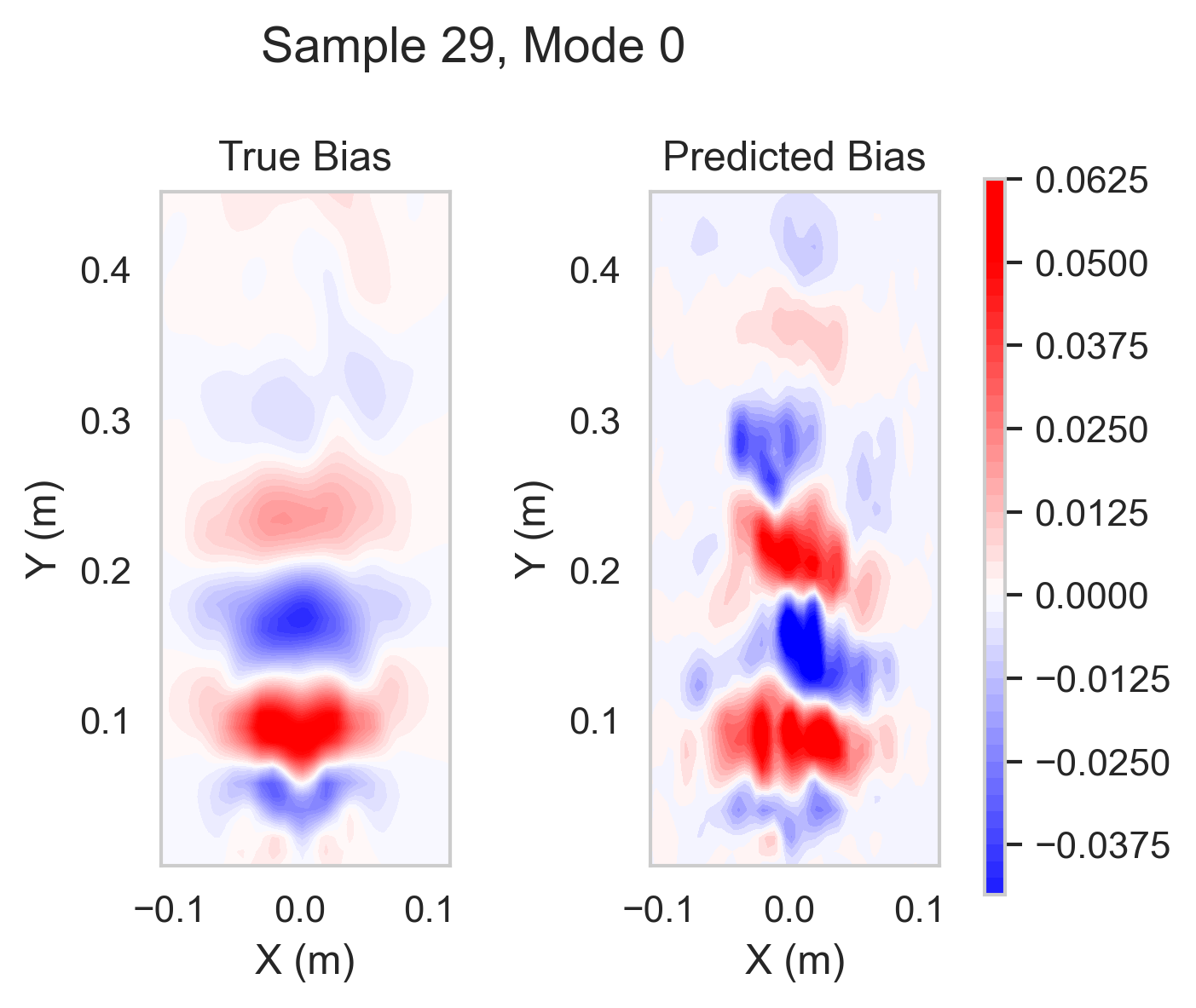}
    }
    \caption{ Performance of structure predictor. The first row is the temporal coefficients of the POD modes. The second and third rows are the true and predicted structures, respectively}
    \label{fig: m2 results}
\end{figure}

In this study, we set the number of filters in the autoencoder layers to 128, and in the convolutional network, it was set to 64 (i.e., $N_{f1}=128$ and $N_{f2}=64$ in Figure~\ref{fig: CNN interpolation}).
Three neighboring samples used to perform nonlinear interpolations ($n=3$ in Figure~\ref{fig: CNN interpolation}). The learning rate was set to 0.001, and the batch size was set to 32. The Adam optimizer was used to minimize the mean squared error (MSE) loss function.
%
Figure~\ref{fig: m2 results} shows the true bias and the predicted bias. The true bias is computed by subtracting the true spatial mode from the reference spatial mode obtained from the structure predictor, as outlined Eq. \ref{eq:bias computation}. On the other hand, the predicted bias is calculated by the structure corrector.

The results demonstrate that the structure shifts and deformations are successfully captured by the structure corrector.


 In the case of oscillatory asymmetry (sample 20), the corrector successfully captures the tilted oscillatory behavior. However, it is important to note that cases without oscillatory asymmetry (sample 21, 4, 29) are also affected by flow asymmetry, causing the machine to potentially perceive a possibility of slight tilting. Further investigations on detecting the flow asymmetry can be conducted to improve the performance of the corrector. Nevertheless, the structure corrector performs well in correcting regions with significant bias overall.

%

\begin{figure}
    \centering
    \subfloat[Sample 20]{
        \includegraphics[width = 0.5\textwidth]{ 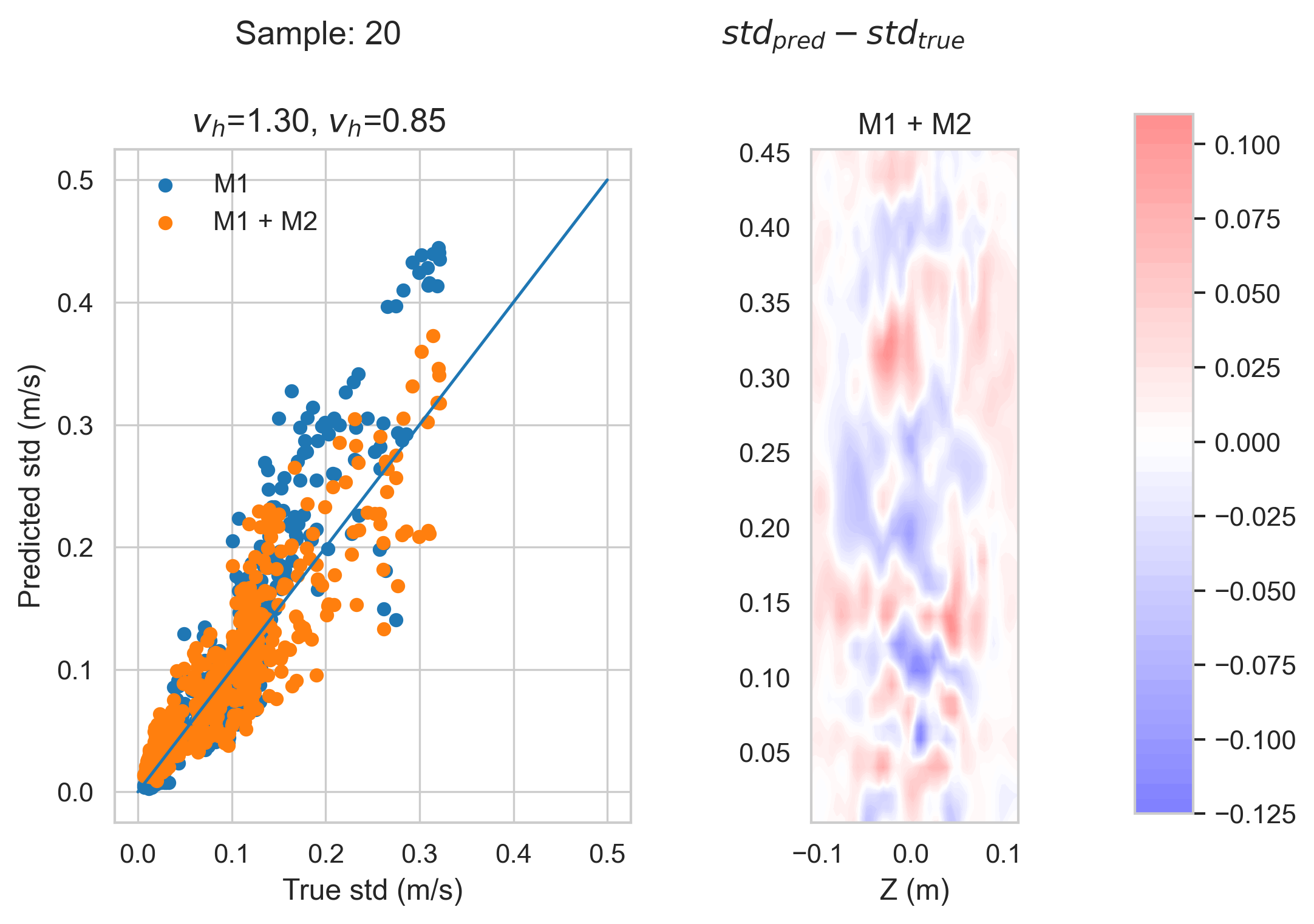}
    }
    \subfloat[Sample 21]{
        \includegraphics[width = 0.5\textwidth]{ 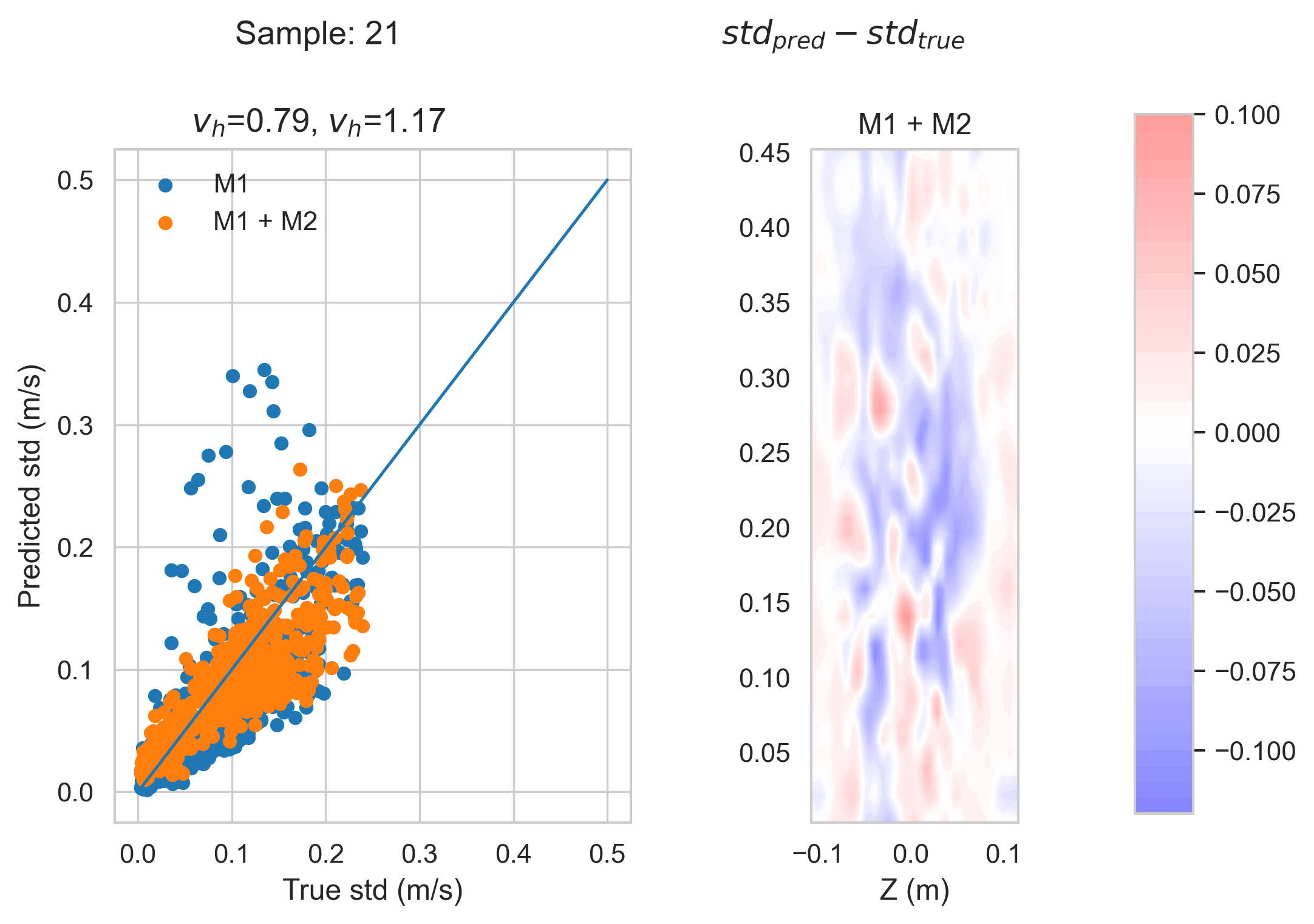}
    }\\
    \subfloat[Sample 4]{
        \includegraphics[width = 0.5\textwidth]{ 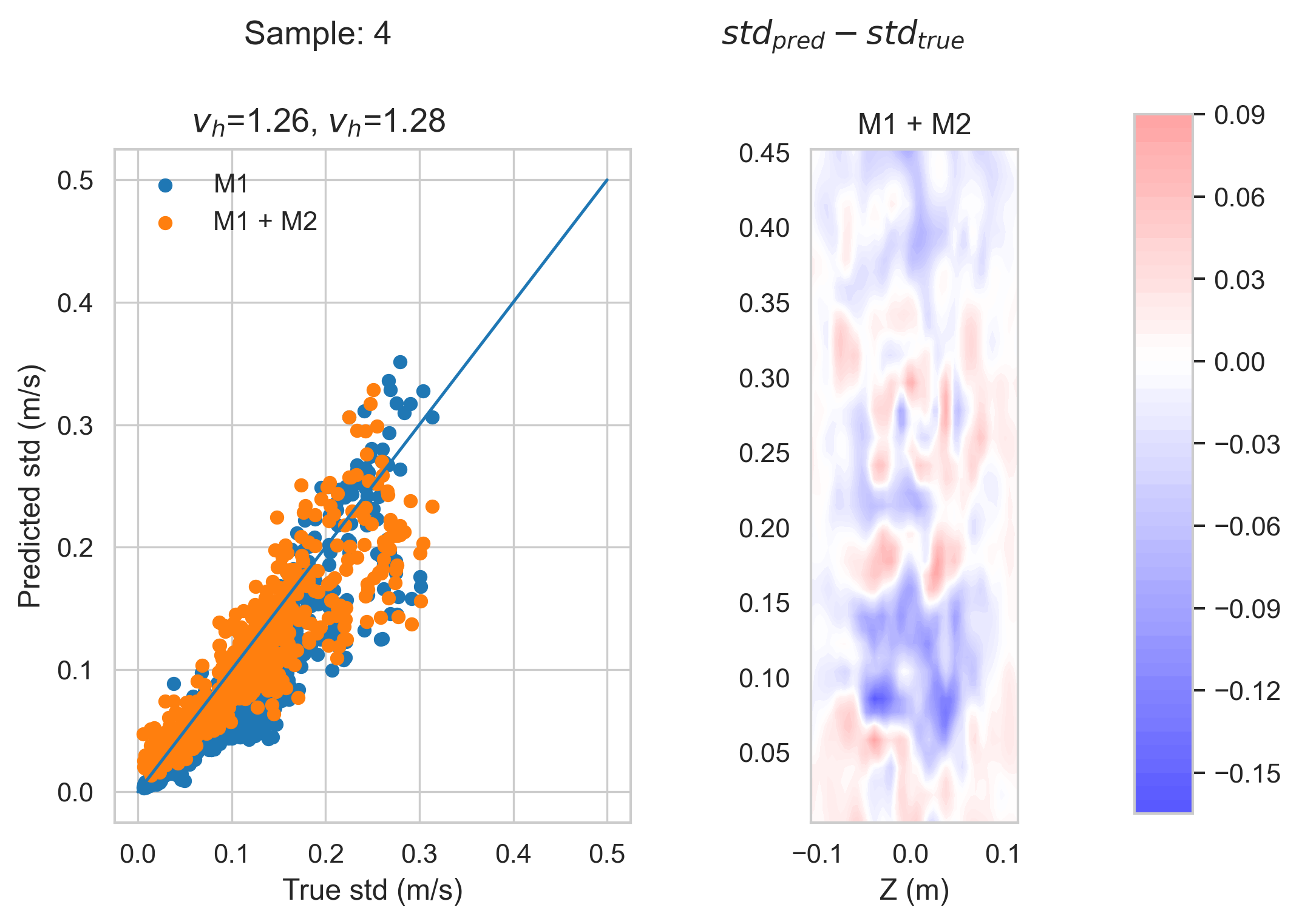}
    }
    \subfloat[Sample 29]{
        \includegraphics[width = 0.5\textwidth]{ 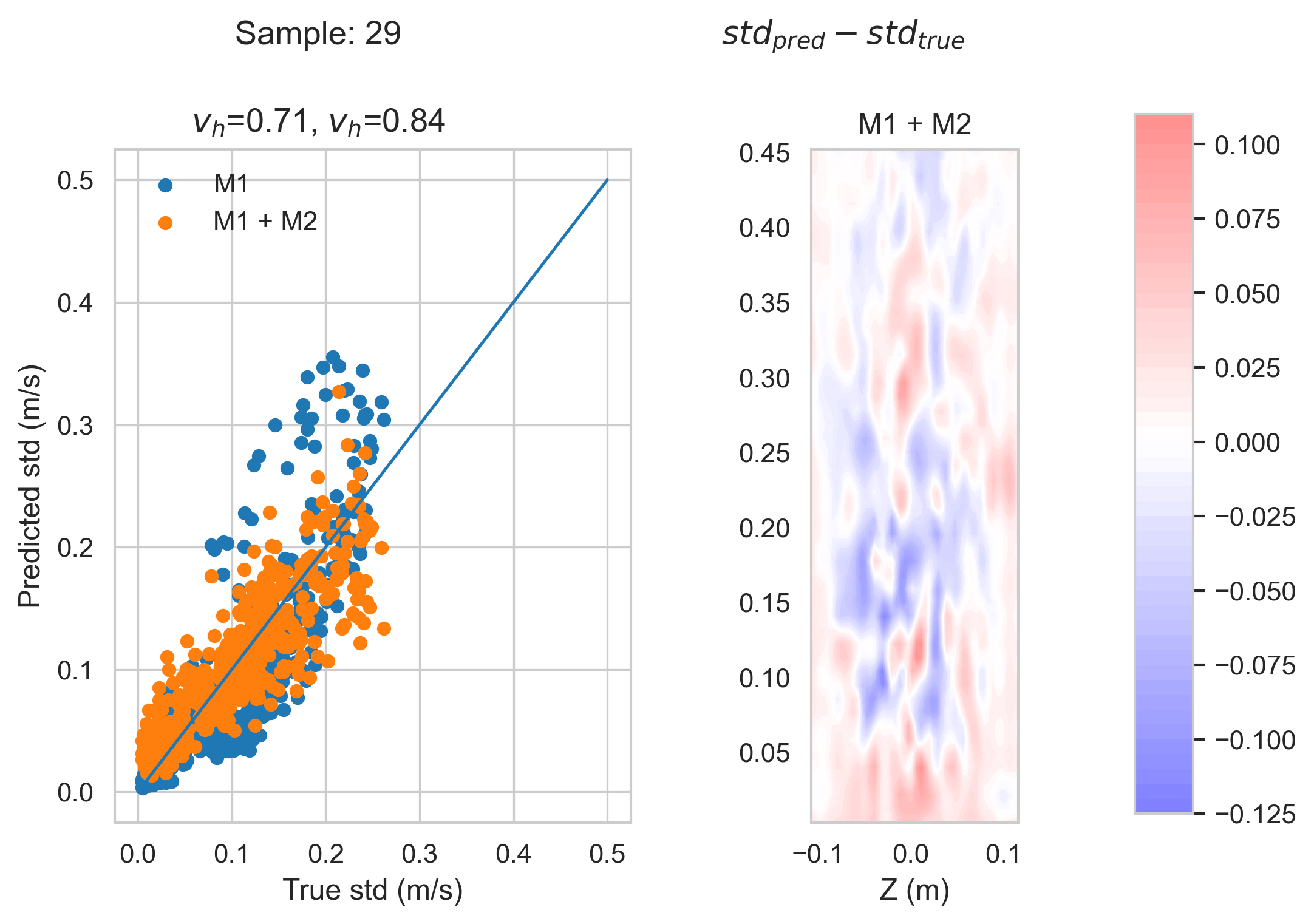}
    }
    \caption{Comparison of standard deviations of the field. The term "M1" indicates the results without bias correction (machine 1 only), and "M1+M2" represents the results with the bias corrector (machine 1 and machine 2)}
    \label{fig: m2 STD}
\end{figure}
\subsection{Spatiotemporal Field Emulation}

The standard deviations of the spatial-temporal fluctuation are show in Figure~\ref{fig: m2 STD}. Here "M1" indicates the results without bias correction (machine 1 only), and "M1+M2" represents the results with the bias corrector (machine 1 and machine 2).  It can be seen that the corrector largely reduces the prediction error and corrects the nonlinear bias in the domain. 

The total root-mean-square error (RMSE) is $0.164 (m/s)$.  
Figure~\ref{fig: m2 STD} shows the comparison between the true and predicted standard deviations of the four testing cases. The total root-mean-square error (RMSE) is $0.164 (m/s)$ for all 2304 spatial points. 
The result shows that the proposed framework has great potential for learning the spatial-temporal field.  

\begin{figure}
    \centering
    \includegraphics[width=\textwidth]{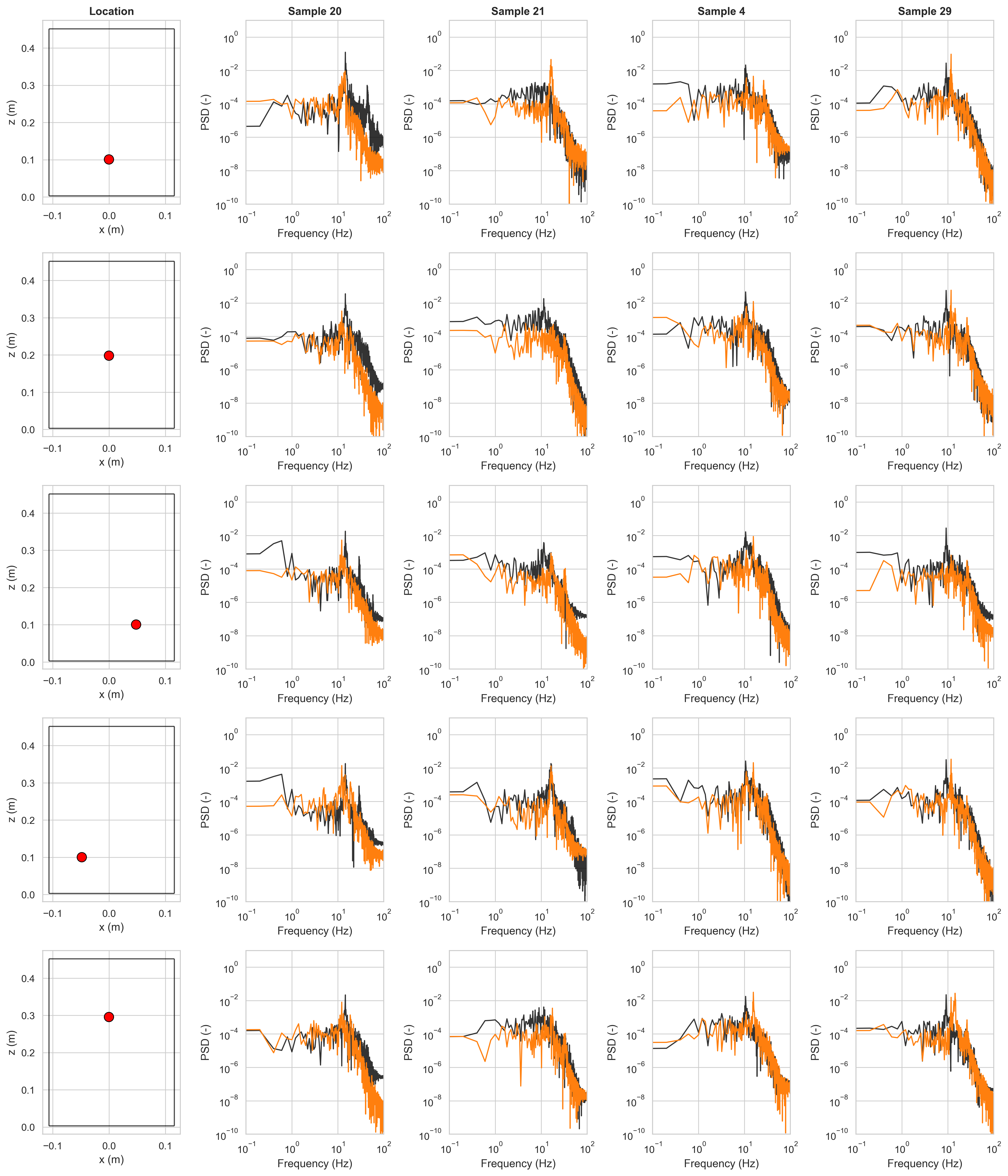}
    \caption{Comparison of the power density spectra of the true and predicted fields at different locations. Black: true; Orange: emulated}
    \label{fig: probes}
\end{figure}

The result of signal emulations at different probes are also shown in  Figure~\ref{fig: probes}. 

The power distributions are well captured by the machines. The PSDs are generally well-predicted in the trend, and the low-frequency peaks are well predicted by the machine. In the context of thermal striping, the low frequency signals $(f=10^{-1} -  10^1 Hz)$ are the critical frequencies that pose a concern in high-cycle thermal fatigue, as the frequencies in this range often lead to the maximum peak-to-valley stress ranges and the frequencies beyond this range can be attenuated by the thermal inertial of the solid structures\cite{smithReportOECDNEAVattenfall2011}. 

The reduction of computational time of the proposed framework is shown in Table~\ref{table: reduction in time}. The trained framework can accelerate the runtime up to 7 orders of magnitude compared to the full-order high-fidelity model. This demonstrates the potential advantage of using this framework for design evaluation, operational forecasting, and online monitoring. 

\begin{table}
    \centering
    \begin{tabularx}{\textwidth}{ccX}
        \toprule
        {} & {Training time} & {Runtime} \\
        \midrule
        {Full model} & {4,140 hr\footnote{}} & {165.6 hr \footnote{one full-order simulation}} \\
        \multirow{2}{*}{\centering proposed framework} & {structure predictor: $21.2(s)$} & \multirow{2}{*}{\centering 38 ($\mu s$)} \\
        {} & {structure corrector: 21 (min)} & { } \\
        \midrule
        \multicolumn{2}{c}{Runtime acceleration (Full-order model/proposed framework)} & {$1.57\times 10^7$}\\
        \bottomrule
    \end{tabularx}
    \caption{The computational time for the full-order CFD simulation (running on a single processor) and framework surrogate}
    \label{table: reduction in time}
\end{table}

\section{Conclusion}
\label{sec:conclusion}
Spatiotemporal prediction of thermal striping is a challenging problem as it requires very high computational cost to resolve the complex turbulence structures interactions at a wide range of scales. For multi-query problems such as design optimization or uncertainty quantification, there is a strong need to leverage a limited number of high-resolution simulations to explore the design space effectively. This work has proposed a two-level machine learning framework for spatial-temporal field emulation of flow and thermal oscillations in industrial applications. The framework is based on the library-lookup-and-correction technique. A structured library containing the local low-dimensional manifolds is first constructed; a structure predictor is designed to down-select the most probable structures from the library. A structure corrector is then trained to reduce the bias between the library and true values.  

Proper orthogonal decomposition (POD) is chosen as the basis construction technique due to the physical relevance of the flow structures identified. Instead of performing POD in the global parametric domain, the present work adopts a local strategy in which POD is carried out on the selected samples in the parameter space. This method increases the reducibility and the efficiency of systems showing high parameter variability. Machine learning algorithms are used for structure lookup and bias correction. The method is non-intrusive, making it highly versatile for different industrial applications. An industrial application, triple jet striping, is selected to demonstrate the method's ability to predict spatial-temporal signals. We show that the spatial-temporal field prediction can be well achieved in a highly nonlinear experiment. 

The current framework is expected to be applicable to other thermal striping applications, such as turbulence mixing in T-junctions and similar physics phenomena that are driven by large coherent structures moving in the flow. Future work involves integrating the proposed framework with an adaptive sampling/space-filling method to reduce training time and improve efficiency. 

\section*{Acknowledgement}
The research is support by the Generating Electricity Managed by Intelligent Nuclear Access (GEMINA) program of the U.S. Advanced Research Project Agency - Energy (ARPA-E) under Award Number [DE-AR0001295]. The support is gratefully appreciated.

\section*{References}

\bibliography{MyLibrary}

\begin{thebibliography}{10}
\expandafter\ifx\csname url\endcsname\relax
  \def\url#1{\texttt{#1}}\fi
\expandafter\ifx\csname urlprefix\endcsname\relax\def\urlprefix{URL }\fi
\expandafter\ifx\csname href\endcsname\relax
  \def\href#1#2{#2} \def\path#1{#1}\fi

\bibitem{bennerModelReductionApproximation2017}
P.~Benner, M.~Ohlberger, A.~Cohen, K.~Willcox, Model Reduction and Approximation: Theory and Algorithms, {SIAM}, 2017.

\bibitem{heSupervisedMachineLearning2019}
L.~He, D.~K. Tafti, A supervised machine learning approach for predicting variable drag forces on spherical particles in suspension, Powder Technology 345 (2019) 379--389.
\newblock \href {https://doi.org/10.1016/j.powtec.2019.01.013} {\path{doi:10.1016/j.powtec.2019.01.013}}.

\bibitem{jaffarPredictionDragForce2020}
F.~Jaffar, T.~Farid, M.~Sajid, Y.~Ayaz, M.~J. Khan, Prediction of {{Drag Force}} on {{Vehicles}} in a {{Platoon Configuration Using Machine Learning}}, IEEE Access 8 (2020) 201823--201834.
\newblock \href {https://doi.org/10.1109/ACCESS.2020.3035318} {\path{doi:10.1109/ACCESS.2020.3035318}}.

\bibitem{wuPredictingInterfacialThermal2019}
Y.-J. Wu, L.~Fang, Y.~Xu, Predicting interfacial thermal resistance by machine learning, npj Computational Materials 5~(1) (2019) 1--8.
\newblock \href {https://doi.org/10.1038/s41524-019-0193-0} {\path{doi:10.1038/s41524-019-0193-0}}.

\bibitem{parkWallTemperaturePrediction2020}
H.~M. Park, J.~H. Lee, K.~D. Kim, Wall temperature prediction at critical heat flux using a machine learning model, Annals of Nuclear Energy 141 (2020) 107334.
\newblock \href {https://doi.org/10.1016/j.anucene.2020.107334} {\path{doi:10.1016/j.anucene.2020.107334}}.

\bibitem{zhaoPredictionCriticalHeat2020}
X.~Zhao, K.~Shirvan, R.~K. Salko, F.~Guo, On the prediction of critical heat flux using a physics-informed machine learning-aided framework, Applied Thermal Engineering 164 (2020) 114540.
\newblock \href {https://doi.org/10.1016/j.applthermaleng.2019.114540} {\path{doi:10.1016/j.applthermaleng.2019.114540}}.

\bibitem{jiangInterpretableFrameworkDatadriven2021}
C.~Jiang, R.~Vinuesa, R.~Chen, J.~Mi, S.~Laima, H.~Li, An interpretable framework of data-driven turbulence modeling using deep neural networks, Physics of Fluids 33~(5) (2021) 055133.
\newblock \href {https://doi.org/10.1063/5.0048909} {\path{doi:10.1063/5.0048909}}.

\bibitem{lingReynoldsAveragedTurbulence2016}
J.~Ling, A.~Kurzawski, J.~Templeton, Reynolds averaged turbulence modelling using deep neural networks with embedded invariance, Journal of Fluid Mechanics 807 (2016) 155--166.
\newblock \href {https://doi.org/10.1017/jfm.2016.615} {\path{doi:10.1017/jfm.2016.615}}.

\bibitem{wangPhysicsinformedMachineLearning2017}
J.-X. Wang, J.-L. Wu, H.~Xiao, Physics-informed machine learning approach for reconstructing {{Reynolds}} stress modeling discrepancies based on {{DNS}} data, Physical Review Fluids 2~(3) (2017) 034603.
\newblock \href {https://doi.org/10.1103/PhysRevFluids.2.034603} {\path{doi:10.1103/PhysRevFluids.2.034603}}.

\bibitem{bruntonMachineLearningFluid2020}
S.~L. Brunton, B.~R. Noack, P.~Koumoutsakos, Machine {{Learning}} for {{Fluid Mechanics}}, Annual Review of Fluid Mechanics 52~(1) (2020) 477--508.
\newblock \href {https://doi.org/10.1146/annurev-fluid-010719-060214} {\path{doi:10.1146/annurev-fluid-010719-060214}}.

\bibitem{changClassificationMachineLearning2019}
C.-W. Chang, N.~T. Dinh, Classification of machine learning frameworks for data-driven thermal fluid models, International Journal of Thermal Sciences 135 (2019) 559--579.
\newblock \href {https://doi.org/10.1016/j.ijthermalsci.2018.09.002} {\path{doi:10.1016/j.ijthermalsci.2018.09.002}}.

\bibitem{chenGreedyNonintrusiveReduced2018}
W.~Chen, J.~S. Hesthaven, B.~Junqiang, Y.~Qiu, Z.~Yang, Y.~Tihao, Greedy nonintrusive reduced order model for fluid dynamics, AIAA Journal 56~(12) (2018) 4927--4943.

\bibitem{xiaoNonintrusiveReducedOrder2015}
D.~Xiao, F.~Fang, A.~G. Buchan, C.~C. Pain, I.~M. Navon, A.~Muggeridge, Non-intrusive reduced order modelling of the {{Navier}}\textendash{{Stokes}} equations, Computer Methods in Applied Mechanics and Engineering 293 (2015) 522--541.
\newblock \href {https://doi.org/10.1016/j.cma.2015.05.015} {\path{doi:10.1016/j.cma.2015.05.015}}.

\bibitem{xiaoNonintrusiveReducedorderModelling2015}
D.~Xiao, F.~Fang, C.~Pain, G.~Hu, Non-intrusive reduced-order modelling of the {{Navier}}\textendash{{Stokes}} equations based on {{RBF}} interpolation, International Journal for Numerical Methods in Fluids 79~(11) (2015) 580--595.
\newblock \href {https://doi.org/10.1002/fld.4066} {\path{doi:10.1002/fld.4066}}.

\bibitem{yuNonintrusiveReducedorderModeling2019}
J.~Yu, C.~Yan, M.~Guo, Non-intrusive reduced-order modeling for fluid problems: {{A}} brief review, Proceedings of the Institution of Mechanical Engineers, Part G: Journal of Aerospace Engineering 233~(16) (2019) 5896--5912.

\bibitem{gonzalezDeepConvolutionalRecurrent2018}
F.~J. Gonzalez, M.~Balajewicz, Deep convolutional recurrent autoencoders for learning low-dimensional feature dynamics of fluid systems (Aug. 2018).
\newblock \href {http://arxiv.org/abs/1808.01346} {\path{arXiv:1808.01346}}, \href {https://doi.org/10.48550/arXiv.1808.01346} {\path{doi:10.48550/arXiv.1808.01346}}.

\bibitem{leeModelReductionDynamical2020}
K.~Lee, K.~T. Carlberg, Model reduction of dynamical systems on nonlinear manifolds using deep convolutional autoencoders, Journal of Computational Physics 404 (2020) 108973.
\newblock \href {https://doi.org/10.1016/j.jcp.2019.108973} {\path{doi:10.1016/j.jcp.2019.108973}}.

\bibitem{shamsSynthesisCFDBenchmarking2018}
A.~Shams, N.~Edh, K.~Angele, P.~Veber, R.~Howard, O.~Braillard, S.~Chapuliot, E.~Severac, E.~Karabaki, J.~Seichter, B.~Niceno, Synthesis of a {{CFD}} benchmarking exercise for a {{T-junction}} with wall, Nuclear Engineering and Design 330 (2018) 199--216.
\newblock \href {https://doi.org/10.1016/j.nucengdes.2018.01.049} {\path{doi:10.1016/j.nucengdes.2018.01.049}}.

\bibitem{fengDemonstrationSTRUCTTurbulence2020}
J.~Feng, E.~Baglietto, K.~Tanimoto, Y.~Kondo, Demonstration of the {{STRUCT}} turbulence model for mesh consistent resolution of unsteady thermal mixing in a {{T-junction}}, Nuclear Engineering and Design 361 (2020) 110572.
\newblock \href {https://doi.org/10.1016/j.nucengdes.2020.110572} {\path{doi:10.1016/j.nucengdes.2020.110572}}.

\bibitem{wangHighFidelityDigital2022}
Y.-J. Wang, K.~Shirvan, E.~Baglietto, P.~Tsilifis, A.~Amer, G.~Khan, L.~Wang, High {{Fidelity Digital Twin}} for {{Critical System Assessments}}, in: 19th {{International Topical Meeting}} on {{Nuclear Reactor Thermal Hydraulics}}, {{NURETH-19}}, {Brussels, Belgium}, 2022.

\bibitem{dahlbergDevelopmentEuropeanProcedure2007}
M.~Dahlberg, {\relax KF}.~Nilsson, N.~Taylor, C.~Faidy, U.~Wilke, S.~Chapuliot, D.~Kalkhof, I.~Bretherton, M.~Church, J.~Solin, Development of a {{European}} procedure for assessment of high cycle thermal fatigue in light water reactors: Final report of the {{NESC-thermal}} fatigue project, Tech. Rep. EUR 22763 EN, {NESC} (2007).

\bibitem{kamayaAssessmentThermalFatigue2014}
M.~Kamaya, Assessment of thermal fatigue damage caused by local fluid temperature fluctuation (part {{II}}: Crack growth under thermal stress), Nuclear Engineering and Design 268 (2014) 139--150.
\newblock \href {https://doi.org/10.1016/j.nucengdes.2013.12.042} {\path{doi:10.1016/j.nucengdes.2013.12.042}}.

\bibitem{blondetHighCycleThermal2009}
E.~Blondet, C.~Faidy, High {{Cycle Thermal Fatigue}} in {{French PWR}}, in: 10th {{International Conference}} on {{Nuclear Engineering}}, {American Society of Mechanical Engineers Digital Collection}, 2009, pp. 429--436.
\newblock \href {https://doi.org/10.1115/ICONE10-22762} {\path{doi:10.1115/ICONE10-22762}}.

\bibitem{bonnetReviewCoherentStructures2001}
J.-P. Bonnet, J.~Delville, Review of {{Coherent Structures}} in {{Turbulent Free Shear Flows}} and {{Their Possible Influence}} on {{Computational Methods}}, Flow, Turbulence and Combustion 66~(4) (2001) 333--353.
\newblock \href {https://doi.org/10.1023/A:1013518716755} {\path{doi:10.1023/A:1013518716755}}.

\bibitem{cantwellOrganizedMotionTurbulent1981}
B.~J. Cantwell, Organized {{Motion}} in {{Turbulent Flow}}, Annual Review of Fluid Mechanics 13~(1) (1981) 457--515.
\newblock \href {https://doi.org/10.1146/annurev.fl.13.010181.002325} {\path{doi:10.1146/annurev.fl.13.010181.002325}}.

\bibitem{fiedlerCoherentStructuresTurbulent1988}
H.~E. Fiedler, Coherent structures in turbulent flows, Progress in Aerospace Sciences 25~(3) (1988) 231--269.
\newblock \href {https://doi.org/10.1016/0376-0421(88)90001-2} {\path{doi:10.1016/0376-0421(88)90001-2}}.

\bibitem{popeTurbulentFlows2000}
S.~B. Pope, Turbulent Flows, {Cambridge university press}, 2000.

\bibitem{lumleyStructureInhomogeneousTurbulent1967}
J.~L. Lumley, The structure of inhomogeneous turbulent flows, Atmospheric turbulence and radio wave propagation (1967).

\bibitem{lumleyCoherentStructuresTurbulence1981}
J.~L. Lumley, Coherent {{Structures}} in {{Turbulence}}, in: R.~E. Meyer (Ed.), Transition and {{Turbulence}}, {Academic Press}, 1981, pp. 215--242.
\newblock \href {https://doi.org/10.1016/B978-0-12-493240-1.50017-X} {\path{doi:10.1016/B978-0-12-493240-1.50017-X}}.

\bibitem{bidanFilmCoolingJetsAnalyzed2013}
G.~Bidan, D.~Nikitopoulos, Film-{{Cooling Jets Analyzed}} with {{Proper Orthogonal Decomposition}} and {{Dynamic Mode Decomposition}}, in: 43rd {{Fluid Dynamics Conference}}, {American Institute of Aeronautics and Astronautics}, {San Diego, CA}, 2013.
\newblock \href {https://doi.org/10.2514/6.2013-2970} {\path{doi:10.2514/6.2013-2970}}.

\bibitem{kalpaklivesterPODAnalysisTurbulent2015}
A.~Kalpakli~Vester, R.~{\"O}rl{\"u}, P.~H. Alfredsson, {{POD}} analysis of the turbulent flow downstream a mild and sharp bend, Experiments in Fluids 56~(3) (2015) 57.
\newblock \href {https://doi.org/10.1007/s00348-015-1926-6} {\path{doi:10.1007/s00348-015-1926-6}}.

\bibitem{meyerTurbulentJetCrossflow2007}
K.~E. Meyer, J.~M. Pedersen, O.~{\"O}zcan, A turbulent jet in crossflow analysed with proper orthogonal decomposition, Journal of Fluid Mechanics 583 (2007) 199--227.
\newblock \href {https://doi.org/10.1017/S0022112007006143} {\path{doi:10.1017/S0022112007006143}}.

\bibitem{wuProperOrthogonalDecomposition2019}
Z.~Wu, D.~Laurence, S.~Utyuzhnikov, I.~Afgan, Proper orthogonal decomposition and dynamic mode decomposition of jet in channel crossflow, Nuclear Engineering and Design 344 (2019) 54--68.
\newblock \href {https://doi.org/10.1016/j.nucengdes.2019.01.015} {\path{doi:10.1016/j.nucengdes.2019.01.015}}.

\bibitem{gongVideoFrameInterpolation}
Z.~Gong, S.~Mall, Video {{Frame Interpolation}} and {{Extrapolation}}.

\bibitem{tokuhiroExperimentalInvestigationThermal1999}
A.~Tokuhiro, N.~Kimura, An experimental investigation on thermal striping: {{Mixing}} phenomena of a vertical non-buoyant jet with two adjacent buoyant jets as measured by ultrasound {{Doppler}} velocimetry, Nuclear Engineering and Design 188~(1) (1999) 49--73.
\newblock \href {https://doi.org/10.1016/S0029-5493(99)00006-0} {\path{doi:10.1016/S0029-5493(99)00006-0}}.

\bibitem{xuSecondGenerationURANS2020a}
L.~Xu, A second generation {{URANS}} approach for application to aerodynamic design and optimization in the automotive industry, Ph.D. thesis, Massachusetts Institute of Technology (2020).

\bibitem{actonStructureBasedResolutionTurbulence2015}
M.~J. Acton, G.~Lenci, E.~Baglietto, Structure-{{Based Resolution}} of {{Turbulence}} for {{Sodium Fast Reactor Thermal Striping Applications}}, in: Proceeding of the {{International Topical Meeting}} on {{Nuclear Reactor Thermal Hydraulics}}, 2015.

\bibitem{fengSTRUCTurebasedURANSSimulations2018}
J.~Feng, T.~Frahi, E.~Baglietto, {{STRUCTure-based URANS}} simulations of thermal mixing in {{T-junctions}}, Nuclear Engineering and Design 340 (2018) 275--299.
\newblock \href {https://doi.org/10.1016/j.nucengdes.2018.10.002} {\path{doi:10.1016/j.nucengdes.2018.10.002}}.

\bibitem{simcenterDocumentationVersion20192019}
S.-C. Simcenter, Documentation {{Version}} 2019.1, Siemens PLM Software (2019).

\bibitem{kimuraExperimentalInvestigationTransfer2007}
N.~Kimura, H.~Miyakoshi, {Hiroyuki Miyakoshi}, {Hideki Kamide}, H.~Kamide, Experimental investigation on transfer characteristics of temperature fluctuation from liquid sodium to wall in parallel triple-jet, International Journal of Heat and Mass Transfer 50~(9) (2007) 2024--2036.
\newblock \href {https://doi.org/10.1016/j.ijheatmasstransfer.2006.09.030} {\path{doi:10.1016/j.ijheatmasstransfer.2006.09.030}}.

\bibitem{smithReportOECDNEAVattenfall2011}
{\relax BL}.~Smith, {\relax JH}.~Mahaffy, K.~Angele, J.~Westin, Report of the {{OECD}}/{{NEA-Vattenfall T-junction Benchmark}} exercise, Tech. rep. (2011).

\end{thebibliography}

\end{document}